\documentclass[journal]{IEEEtran}

\usepackage{soul}
\usepackage{subfigure}
\usepackage{amsmath}
\usepackage{mathtools,amssymb}
\usepackage{subfigure}
\usepackage{graphicx}
\usepackage{color}
\usepackage{dblfloatfix}
\usepackage[framemethod=tikz]{mdframed}

\usepackage{algorithm}
\usepackage{algpseudocode}
\algrenewcommand\algorithmicrequire{\textbf{Initialize:}}

\newcommand{\mbf}[1]{\mathbf{#1} }

\newcommand{\X}{{\cal{X}}}
\renewcommand{\H}{{\cal{H}}}

\newcommand{\E}{{\mathbb{E}}}

\newcommand{\fig}[1]{Fig.~\ref{fig:#1}}
\renewcommand{\sec}[1]{Section~\ref{sec:#1}}
\newcommand{\eq}[1]{(\ref{eq:#1})}

\ifCLASSINFOpdf
\else
\fi

\begin{document}
%
\markboth{Journal of \LaTeX\ Class Files,~Vol.~14, No.~8, August~2015}%
{Shell \MakeLowercase{\textit{et al.}}: Bare Demo of IEEEtran.cls for IEEE Journals}
\title{BICM-compatible Rate Adaptive Geometric Constellation Shaping Using Optimized Many-to-one Labeling}

\author{Metodi Plamenov Yankov,~\IEEEmembership{Senior Member,~IEEE, Member,~OPTICA}, Smaranika Swain,~\IEEEmembership{Member,~OPTICA}, Ognjen Jovanovic, ~\IEEEmembership{Member,~IEEE}, Darko Zibar,~\IEEEmembership{Member,~OPTICA}, and Francesco Da Ros,~\IEEEmembership{Senior Member,~OPTICA, Senior Member,~IEEE}%
\thanks{Metodi Plamenov Yankov, Smaranika Swain, Darko Zibar and Francesco Da Ros are with the Department of Electrical and Photonics and Electrical Engineering, Technical University of Denmark, 2800 Kgs. Lyngby, Denmark. Ognjen Jovanovic is with AdTran Networks SE, 82152 Martinsried/Munich, Germany. e-mail: meya@dtu.dk}}

\maketitle

\begin{abstract}
In this paper, a rate adaptive geometric constellaion shaping (GCS) scheme which is fully backward-compatible with existing state of the art bit-interleaved coded modulation (BICM) systems is proposed and experimentally demonstrated. The system relies on optimization of the positions of the quadrature amplitude modulation (QAM) points on the I/Q plane for maximized achievable information rate, while maintaining quantization and fiber nonlinear noise robustness. Furthermore, `dummy' bits are multiplexed with coded bits before mapping to symbols. Rate adaptivity is achieved by tuning the ratio of coded and `dummy' bits, while maintaining a fixed forward error-correction block and a fixed modulation format size. The points' positions and their labeling are optimized using automatic differentiation in a many-to-one fashion such that the performance of the `dummy' bits is sacrificed in favor of the performance of the data bits. The proposed GCS scheme is compared to a time-sharing hybrid (TH) QAM modulation and the now mainstream probabilistic amplitude shaping (PAS) scheme employing a Maxwell-Boltzmann probability mass function. The TH without shaping is outperformed for all studied data rates in a simulated linear channel by up to 0.7 dB. In a linear channel, PAS is shown to outperform the proposed GCS scheme, while similar performances are reported for PAS and the proposed GCS in a simulated nonlinear fiber channel. 

The GCS scheme is experimentally demonstrated in a multi-span recirculating loop coherent optical fiber transmission system with a total distance of up to 3000 km. Near-continuous zero-error flexible throughput is reported as a function of the transmission distance. Up to 1-2 spans of increased reach gains are achieved at the same net data rate w.r.t. conventional QAM. At a given distance, up to 0.79 bits/2D symbol of gain w.r.t. conventional QAM is achieved. In the experiment, similar performance to PAS is demonstrated.
\end{abstract}


\section{Introduction}
Over the recent years, coherent fiber optic communications have seen a steady improvement in their operating spectral efficiency \cite{NokiaWP}. This has been enabled by the adoption of high order modulation formats, primarily from the quadrature amplitude modulation (QAM) family with the conventional regular distribution of the points on the I/Q plane. Such constellations are known to exhibit a constellation shaping gap to the Shannon capacity due to their typically uniform probability mass function (PMF) \cite{Cover}. 

Furthermore, in order to minimize the margins and thus improve the efficient usage of the available quality of transmission (QoT), rate adaptivity is imperative in heterogenous networks with a wide QoT distribution of different signal paths. Ageing of optical equipment also requires that the transmission rate follows the QoT trends, which may degrade over the system lifespan. 

For fixed bandwidth transmission, rate adaptivity is typically approached by one of the following options:
\begin{enumerate}
 \item adopting a forward error correction (FEC) code of variable rate;
 \item support flexibility of the modulation format size;
 \item adopt a time-sharing hybrid (TH) transmission, where the modulation format is variable in pre-determined time slots.
\end{enumerate}
The first option (in combination with Option 2)) is viable with moderate speeds and is the current de-facto method for wireless communications. In e.g. the 5G standard for cellular communications, rate-adaptive FEC is achieved by puncturing of the adopted low-density parity check (LDPC) code. Punctured LDPC codes are not as efficient as fixed-rate irregular LDPC codes (as e.g. adopted in the DVBS-2 satellite communications standard), which can be specifically tailored to the target rate. However, the emerging standards of 800G and 1.6T for coherent pluggable modules may prohibit variable-rate FEC due to the associated chip area and complexity requirements. On the other hand, Option 2) alone only provides a rate flexibility with a step of $R$ bits/symbol where $R$ is the rate of the FEC, which is too coarse to cover the wide range of QoT experienced in a heterogenous network. Option 3) is a trade-off, which however may suffer a performance penalty w.r.t. Option 1) \cite[Fig. 11]{Cho}, and is also related to increased complexity digital signal processing. State of the art standardized coherent optical systems are limited to Option 2) (e.g. the oFEC standard for data center interconnects \cite{oFEC}).  

Much effort has been spent in the last decade to reduce the shaping gap by optimizing the PMF \cite{Vassilieva}. This optimization can be approached geometrically (geometric constellation shaping (GCS)) or probabilistically (probabilistic constellation shaping (PCS)). In the former case, the positions of conventional QAM constellation points on the I/Q plane are optimized, typically to achieve an approximation of a Gaussian distribution \cite{IPM}. In the latter, the probability of occurrence of each point is optimized instead, and the points can maintain their regular locations \cite{Cho}. For a linear additive white Gaussian noise (AWGN) channel, a near-optimal PMF of regular QAM constellations was found to be the Maxwell-Boltzmann (MB) \cite{Frank}, also known as the quantized Gaussian distribution. In optical fibers, another key metric used to optimize the constellation can be its effectiveness in generation of nonlinearities due to the Kerr effect, which is detrimental to coherent transmission and is enhanced with Gaussian distributions \cite{Dar}. The typically contradicting requirements of AWGN tolerance and nonlinear interference noise (NLIN) generation give rise to a non-trivial optimization problem, with only approximate solutions available thus far (highly incomplete list may be \cite{Henrik}\cite{Secondini}\cite{Jones2019}\cite{IPM}\cite{MariiaRipple}\cite{Cho}). 

Geometric shaping schemes, while addressing the shaping gap problem, are typically not rate adaptive and require the same approaches as standard QAM listed above to achieve rate adaptivity. On the other hand, PCS, and especially the probabilistic amplitude shaping (PAS) scheme \cite{Bocherer} is emerging as a viable solution to the shaping problem while also addressing the rate adaptivity of the system. While ideal distribution matchers (DMs) for PAS require impractically long lengths which are difficult to realize in a parallelized hardware, practical schemes do exist. Slightly sub-optimal DMs of minimal rate loss and lower implementation complexity have already been developed and are a hot topic of current research \cite{Stella, Yoshida, Yunus} in an effort to maintain rate adaptation capability, albeit at a slightly lower achievable shaping gain.

For future heterogeneous networks of high spectral efficiency and minimal required margin, a low-complexity, efficient, rate adaptive coded modulation system is necessary. This paper addresses this problem by proposing and experimentally demonstrating a GCS system which:
\begin{itemize}
 \item does not require flexible rate FEC;
 \item does not require flexibility in the modulation format size;
 \item does not require a DM;
 \item achieves comparable performance to MB PMF PAS in the presence of nonlinearities;
 \item achieves near-continuous target net data rate.
\end{itemize}

The method was first proposed in \cite{YankovECOC2023, OgnjenCLEO}, and later experimentally demonstrated in \cite{YankovIPC}. This paper extends those works by:
\begin{itemize}
 \item theoretically analyzing the scheme in an AWGN channel scenario;
 \item comparing to a TH modulation;
 \item expanding on the experimental demonstration with various scenarios of interest and detailing the experimental setup;
 \item providing a detailed comparison to PAS;
 \item analyzing in-depth the parametrization of the system;
 \item discussing the potential improvements and future work.
\end{itemize}

The paper is structured as follows. In \sec{method}, the proposed scheme is described and its potential for rate adaptation is shown. In \sec{optimization}, the optimization method which enables the proposed scheme to function in an optimal manner is described. In \sec{AWGN_results}, the proposed system is compared to a TH modulation and PAS in an AWGN channel. In \sec{ideal_results}, the benefits of the proposed method are demonstrated for an idealized model of a coherent optical fiber wavelength division multiplexing (WDM) link. In \sec{setup}, the experimental setup used for demonstration of the proposed scheme is detailed. Furthermore, the procedure for fitting the idealized model to the experimental setup is described. In \sec{exp_results}, the experimental results are shown and a comparison to PAS is made. In \sec{discussion}, additional discussion points are offered. Finally, in \sec{conclusion}, the work is concluded.

\section{Rate adaptivity using many to one mapping}
\label{sec:method}

\begin{figure*}[!t]
\centering
 \includegraphics[trim=0 0 0 0cm, width=1.0\linewidth]{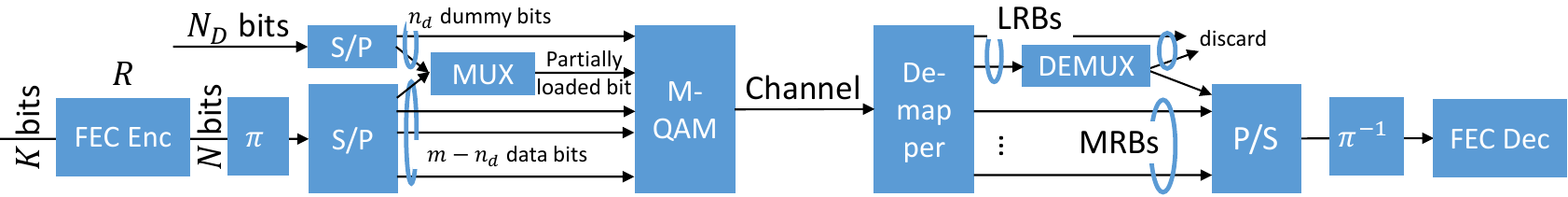}
 \caption{Block diagram of the proposed rate adaptive system.}
 \label{fig:MTO_scheme}
\end{figure*}

The block diagram of the proposed coded modulation (CM) scheme is given in \fig{MTO_scheme}. The data sequence of length $K$ is encoded by an FEC, and then interleaved by interleaver $\pi$. The FEC code has a length of $N$. Each codeword is multiplexed (MUX-ed) with an integer number of pre-defined bits $N_D$ before parallel to serial conversion. The resulting $m$ bit sequences are mapped to the constellation symbols. The constellation is of a size $M=2^m$. The net data rate of transmission is found as $\eta=\frac{K\cdot m}{N+N_D} = R\cdot(m-n_d)$ bits/symbol, where $n_d=\frac{m\cdot N_D}{N+N_D}$ and $R=K/N$ is the rate of the FEC. For fractional $n_d$, the $(m-\lfloor n_d \rfloor)-$th bit of the label, where $\lfloor \cdot \rfloor$ is the rounding down operator, is \textit{partially loaded} and \textit{shared} between data and `dummy' bits. At the receiver, the `dummy' bits are discarded. This means that in the case of soft decision (SD) demapping, their log-likelihood ratios (LLRs) do not need to be calculated. The coded bits are de-interleaved and FEC-decoded.

The `dummy' bits are not used to carry information. During serial to parallel conversion, they are therefore allocated to the least reliable bit (LRB) (in terms of mutual information (MI)) label positions of the constellations in order to maximize the total MI on the most reliable bits (MRBs), which are delivered to the decoder. For binary reflected Gray coded (BRGC) pulse amplitude modulation (PAM), the LRBs are typically the least significant bits (LSBs). 

Because the `dummy' bits do not carry information, constellation symbols which are defined by those label positions do not need to be distinguishable and can be merged to the same location on the I/Q plane. The benefit is increased MI of the other bits. This process may be seen as a many to one mapping (MTOM). In fact, a 64QAM BRGC constellation may be seen as a geometrically shaped 256QAM BRGC constellation with MTOM, where the four points which have identical first six MRBs are merged into a single one. The location of the latter is determined by the average power constraint of the constellation and the restriction that a 64QAM is an outer product of two identical 8PAM constellation with equal spacing between the points. However, the location of the point is in general subject to optimization.

\begin{figure}[!t]
\centering
 \includegraphics[trim=0 0 0 0cm, width=1.0\linewidth]{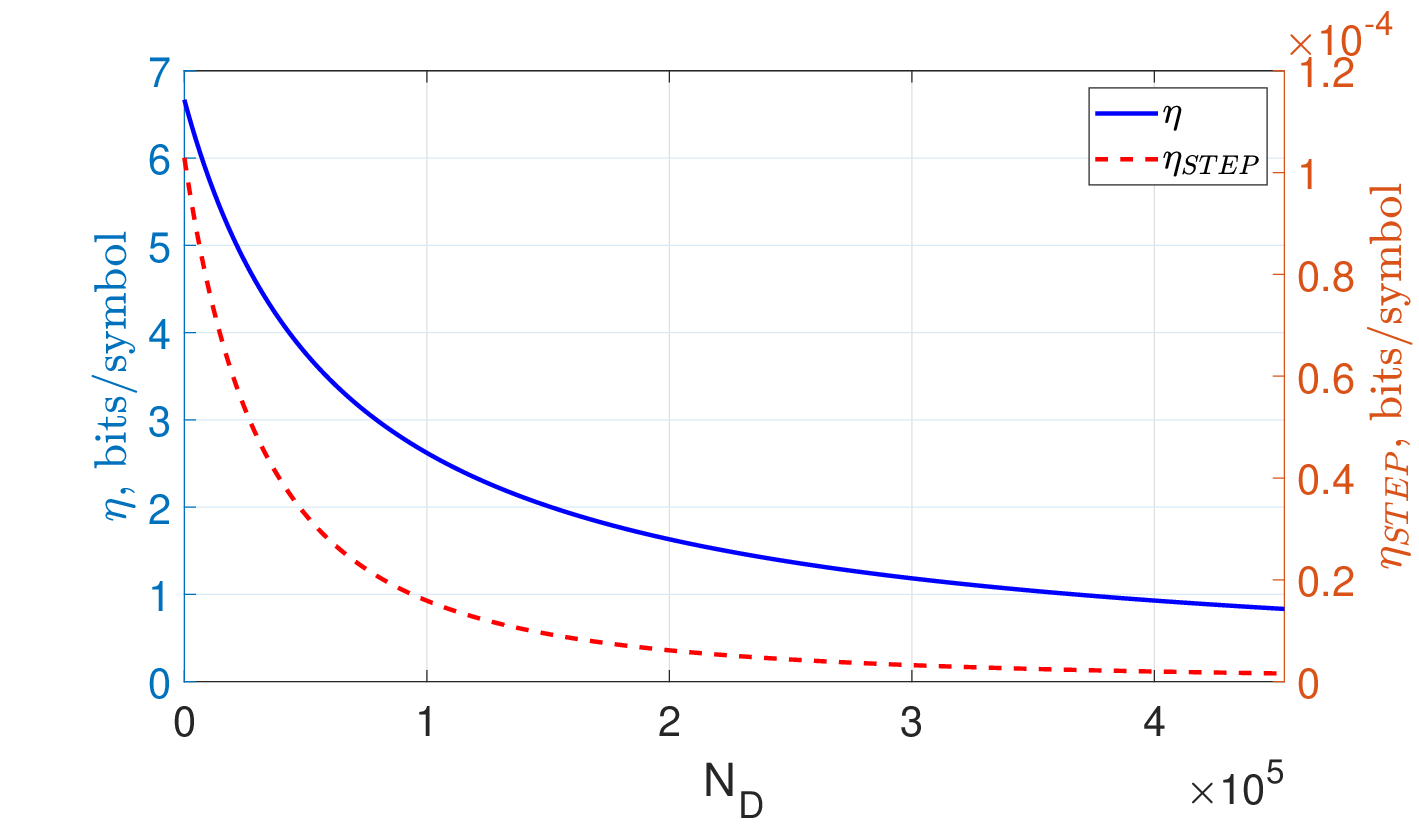}
 \caption{\textbf{(left axis:)} Example target IRs $\eta$ for different choices of the number of dummy bits $N_D$ for $m=8$, $N=64800$ and $K=54000$. \textbf{(right axis:)} minimum change $\eta_{STEP}$ of the target IR.}
 \label{fig:example_rate_step}
\end{figure}

When $N_D=0$, the scheme collapses to conventional bit-interleaved coded modulation (BICM) and full backwards compatibility is therefore achieved. The rate can be tuned in the range $\eta \in \left(0; R\cdot m\right]$ by selecting the number $N_D$ from the range $N_D \in \left[0; \infty\right)$. The rate can be decreased from its maximum value with a (nonlinear) step $\eta_{STEP}$ by incrementing $N_D$ with a step of 1. The possible rates (left axis) and the corresponding steps (right axis) are given in \fig{example_rate_step} for the case of $m=8$, $N=64800$ and $K=54000$, which corresponds to the 20\% FEC overhead DVBS-2 low density parity check (LDPC) code \cite{DVBS2}. The step is below 0.0001 bits/symbol, which for an ideal, capacity-achieving FEC over an additive white Gaussian noise (AWGN) channel is small enough to cover effective signal-to-noise ratio (SNR) changes of $3\cdot 10^{-4}$ dB. The step is larger for shorter FECs, e.g. $\eta_{STEP}<6\cdot10^{-3}$ for $N=1000$, which is still small enough for SNR variations with a step of $3\cdot 10^{-3}$ dB. For comparison, a conventional BICM which can only switch its modulation format size has a rate step of $\eta_{STEP}=R$, which for $R=5/6$ corresponds to SNR variations with a step of $\approx 2.5$ dB\footnote{The supported SNR variations are only approximate and are evaluated by inverting Shannon's formula $\eta=\log_2(1+SNR)$ at a target rate of $\eta=6$ bits/2D symbol and evaluating the SNR difference to the inverted Shannon's formula at $6-\eta_{STEP}$ bits/2D symbol. The true changes are functions of the modulation format, the specific FEC performance and the target rate itself.}.

\section{Optimization method}
\label{sec:optimization}
\subsection{Channel model}
\label{sec:model}
The scheme from \sec{method} is directly applicable to multi-dimensional constellations. In this work, without loss of generality, a 2-dimensional channel is considered, where the two dimensions describe the real and imaginary component of a complex-valued channel model. The constellation symbols $x_k$ at time $k$ are coming from an alphabet $x_k \in \X$ and are therefore complex-valued, i.e. 2-dimensional. Complex-valued AWGN samples $n_k$ are assumed, coming from a Gaussian distribution with a variance $\sigma_{AWGN}^2$. In the case of linear AWGN channel transmission, the variance of the AWGN is given by $\sigma^2=1/SNR$, where unit power constraint is assumed. In the case of fiber optic communications with Kerr nonlinearities, the enhanced Gaussian noise (EGN) model \cite{Carena} can be adopted, according to which the received effective SNR can be modeled as 
\begin{align}
 SNR^{-1} = (SNR_{TRX}^{-1} + SNR_{ASE}^{-1} + SNR_{NLI}^{-1})^{-1}, 
\end{align}
where $SNR_{TRX}$ is the implementation penalty of the transmitter and receiver, $SNR_{ASE}$ is the electrical noise due to amplified spontaneous emission (ASE) amplification noise on the link, and $SNR_{NLI}$ is the in-band noise due to the fiber nonlinearities. All impairments are assumed to be additive, white, and Gaussian. The NLI term is well-described in \cite{Carena, Dar}. Here, we only state for completeness that the variance of the NLIN is a function of the 4-th and 6-th order moment of the input constellation, the cube of the transmission power, as well as the physical fiber parameters (such as dispersion and nonlinear coefficient). The latter are static and independent of the modulation format and coding scheme. The term $SNR_{ASE}$ is calculated based on the noise figure (NF) of the selected amplification scheme \cite{Essiambre}, and the term $SNR_{TRX}$ is predefined and fixed. The variance of the AWGN is therefore $\sigma_{AWGN}^2=SNR_{TRX}^{-1} + SNR_{ASE}^{-1} + SNR_{NLI}^{-1}$. A real-valued quantization noise is generated in each of the real and imaginary dimensions at both the transmitter (corresponding to digital to analog converter (DAC) noise $w_k^{DAC}$) and at the receiver (corresponding to analog to digital converter (ADC) noise $w_k^{ADC}$). Both dimensions of both noise sources are uniformly distributed in the range $\left[ -\Delta/2^{nQbits-1}, \Delta/2^{nQbits-1}\right]$. Here, $nQbits$ is the number of quantization bits, 
\begin{align}
\label{eq:quanRange}
\Delta=1.1 \cdot \max_i \big\{ \max \{ {\mathrm{Re}}\left[\X^i\right], {\mathrm{Im}}\left[\X^i\right] \} \big\}
\end{align}
is the assumed dynamic range of the ADC and DAC, ${\mathrm{Re}}\left[\X^i\right]$ is the real part of the $i-$th point from the constellation set, and ${\mathrm{Im}}\left[\X^i\right]$ is the imaginary part of the $i-$th point from the constellation set. The factor $1.1$ in \eq{quanRange} is arbitrarily chosen to cover the dynamic range of the assumed symmetric constellation and to some extend the additional peaks of the pulse shaped signal. Any non-uniform components of the quantization noise (e.g. due to finite bandwidth with a certain frequency response) are assumed to have a Gaussian distribution and are incorporated in the term $SNR_{TRX}$, because they are assumed to be independent of the constellation format. The memoryless end-to-end channel model is given by 
\begin{align}
 y_k = x_k + n_k + w_k^{DAC} + w_k^{ADC},
\end{align}
where $y_k$ is the channel output.

\subsection{Achievable information rate}
The performance metric of interest which predicts the performance of a SD FEC is the bit-wise MI also known as the generalized mutual information (GMI) in the optical communication community \cite{GMI}. The bit-wise MI is measured in bits/2D symbol. When the signal is corrupted by an AWGN with a known noise variance, the bit-wise MI can be estimated by performing bit-metric de-mapping using the Gaussian likelihood function to extract the symbol likelihoods and then performing marginalization to extract the bit LLRs \cite{tenBrink}. For demapping, a Gaussian likelihood function is used with a variance of 
\begin{align}
\sigma^2 = \sigma_{AWGN}^2 + \left(\frac{\Delta}{2^{nQbits}}\right)^2\cdot\frac{1}{12} 
\end{align}
which is mismatched to the true likelihood of the channel due to the non-uniformity of the quantization noise. This leads to a lower bound on the bit-wise MI. This lower bound will be the achievable information rate (AIR) when applying the same mismatched likelihood for marginalization during transmission. The marginalization itself is performed by using the Gaussian likelihoods together with the fixed, pre-defined labeling function \cite{tenBrink}. The latter is a look-up table (LUT) between integers in the range $\left[1, 2^m\right]$ (each corresponding to a constellation symbol) and a binary label of length $m$. The bit-wise MI is different for each label position, which is the underlying cause of some bits being `more' or `less' reliable. In BICM, due to the typically long interleaving, the average bit-wise MI is a good predictor of the performance of the FEC. 

The AIR of the proposed system when the dummy bits are explicitly discarded (as in \fig{MTO_scheme}) is

\begin{align}
 \label{eq:AIR}
 AIR = & m-\lceil n_d \rceil + \sum_{i=1}^{m-\lceil n_d \rceil} \E_k \left[\log_2 p(\mbf{u}_k^i|y_k) \right] + \notag \\ 
      & + (\lceil n_d \rceil-n_d)\cdot \left(1 + \E_k \left[\log_2 p(\mbf{u}_k^{m-\lfloor n_d \rfloor}|y_k) \right] \right),
\end{align}
where $\lceil \cdot \rceil$ is the rounding up operator, $\mbf{u}_k^i$ is the $i-$th bit in the label $\mbf{u}_k$ of $x_k$ and $p(\mbf{u}_k^i|y_k)$ is the corresponding marginalized posterior probability of that bit. The expectation terms $ \E_k \left[\cdot \right]$ in Eq.~\eq{AIR} correspond to the negative binary cross-entropy (BCE) of the bits on the data-carrying positions in the label. Without loss of generality, the data are assumed to be mapped only to position indexes $\{1, 2, \dots m-\lfloor n_d \rfloor \}$, where the positions are ordered from 1 to $m$ in the order of their reliability (1 being the MRB). The dummy bits are mapped only to positions $\{ m-\lfloor n_d \rfloor, m-\lfloor n_d \rfloor+1, \dots m \}$. The partially-loaded position is $i=m-\lfloor n_d \rfloor$. The first row in Eq.~\eq{AIR} corresponds to the bit-wise MI of the purely data-carrying bit positions in the label, while the second row corresponds to the MI, obtainable from the partially loaded bit.

\subsection{Optimization using automatic differentiation}

\begin{figure*}[!t]
\centering
 \includegraphics[trim=0 0 0 0cm, width=1.0\linewidth]{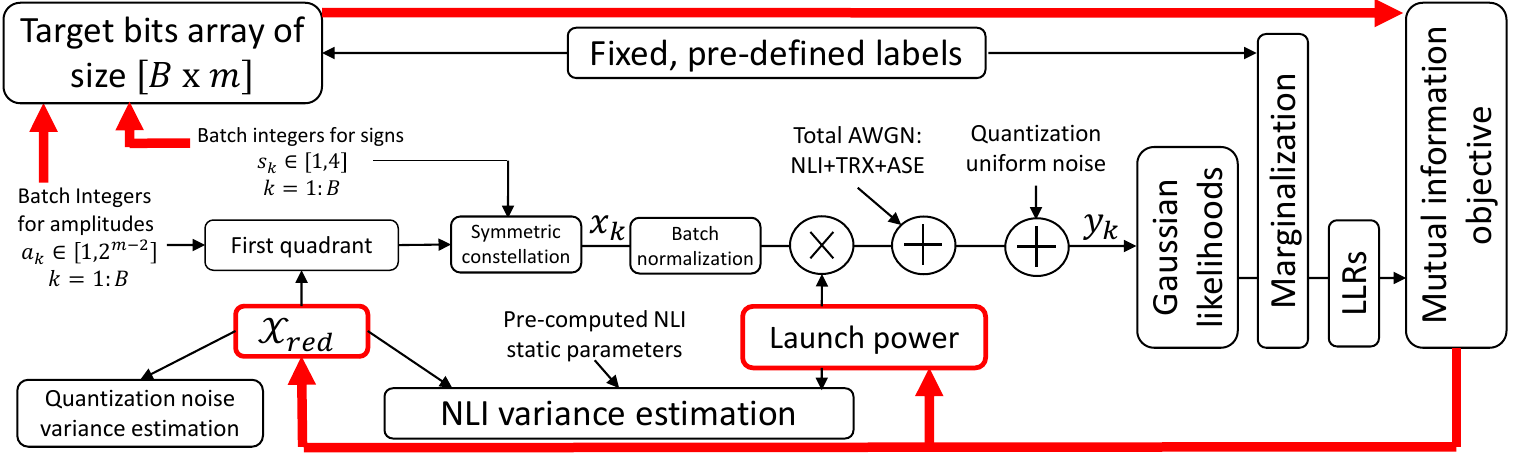}
 \caption{Block diagram of the optimization process. Red, bold arrows indicate the optimization loop.}
 \label{fig:optimization}
\end{figure*}

The bit-wise MI can be optimized w.r.t. the constellation moments and the transmission power using the block diagram in \fig{optimization}. All functions which take place during estimation of the bit-wise MI are differentiable w.r.t. their input, which enables the implementation of automatic differentiation (AD). The method of AD relies on the chain rule for differentiation in order to calculate the gradient of the objective function w.r.t. the desired optimization variables. First, a computational graph is built for the forward model from \sec{model}. The gradients of the target objective function are then sequentially back-propagated through the graph until the optimization variables are reached. Assuming none of the forward model functionalities result in discontinuities in their outputs (which is the case for all functionalities described in \sec{model}), the gradients after full back-propagation are used to update the optimization variables. In the case of classical machine learning of neural network (NN) models, the optimization variables are the weights of the NN. In this paper, the PyTorch AD tool is used for optimization \cite{PyTorch}, where further details of the AD process can also be found.

The process for the optimization of $\X$ and the launch power used in this paper is summarized in Algorithm~\ref{alg:optimization} and can be detailed as follows. 

\begin{algorithm}
    \caption{Optimization algorithm for MTOM GCS transmission.}
    \label{alg:optimization}
    \begin{algorithmic}[1]
    \Require $\X_{red}$ (e.g. 1-st quadrant of regular QAM)
    \Require Integer-to-bit label LUT (e.q. BRGC)
    \Require Fiber parameters ($\gamma$, $D$, NF, span length, number of spans, carrier frequencies)
    \State Compute NLI constants
    \While{not converged} \Comment Main optimization loop
        \State Compute qantization noise variance \Comment Eq.~\eq{quanRange}
        \State Compute NLIN variance \Comment EGN or NLIN model
        \State Generate batch indexes $a_k$, $s_k$
        \State Generate symbols $x_k$
        \State Normalize to launch power
        \State Add AWGN
        \State Add quantization noise
        \State Compute symbol likelihoods
        \State Compute LLRs \Comment using pre-defined labels
        \State Compute objective function \Comment Eq.~\eq{cost}
        \State Compute gradients \Comment using AD
        \State Update $\X_{red}$
        \State Update launch power
    \EndWhile    
    \end{algorithmic}
\end{algorithm}

First, the constellation and labeling function are initialized, e.g. to a conventional BRGC QAM. The LUT labels \textit{do not need to change during optimization}. Random sequence of integers $a_k$ from the range $a_k \in \left[1,2^{m-2}\right]$ are generated of a given length $k = 1,2,\dots B$, where $B$ is known as the \textit{batch size}. This sequence will control the amplitude and phase of the constellation points in the first quadrant of the I/Q plane. The integers correspond to indices of symbols from the constellation $\X_{red}$, which is the subset of $\X$ subject for optimization in this paper. Another sequence of random integers $s_k$ is generated from the range $s_k \in \left[1,2^2\right], k = 1,2,\dots B$, which will control the signs of the real and imaginary part. These signs will be mapped to the two MRBs of the constellation label. Together, these sequences of indices will generate a complex valued symbol sequence batch from the complete constellation $x_k\in\X$. The labeling function is used to generate target bits from the batch symbols. Using the tensor $\X_{red}$, the 4th and 6th order moments are calculated according to the NLIN model. Using the moments, the transmission power and the static fiber parameters, $SNR_{NLI}$ is calculated. 

The intuition behind the simplification of optimizing $\X_{red}$ instead of directly $\X$ is that for phase-symmetric noise, the optimized constellation should also be phase-symmetric. A quadrant-symmetric constellation is an approximation of a phase-symmetric one. This simplification does not only provide a lower-complexity optimization process, but also regularizes the process and helps avoid local optima. For asymmetric channels, the full constellation can be optimized with the same procedure described above. 

In order to optimize the tensor $\X_{red}$ for MTOM transmission instead of standard BICM, the objective function (OF) of interest is
 \begin{align}
 \label{eq:cost}
 OF = m + \sum_{i=1}^{m-\lfloor n_d \rceil} \E_k \left[ \log_2 p(\mbf{u}_k^i|y_k) \right],
 \end{align}
where $\lfloor \cdot \rceil$ is the nearest integer operator (which is discussed below). The objective function disregards the MI of the label positions which mostly carry dummy bits in favor of maximizing the MI of the other bit positions. The effect of this objective function is that points from the constellation which carry identical label on the data-carrying positions will be pushed to the same location because the objective function is effectively not a function of the `dummy' bits. When $n_d=0$, the OF is the AIR of standard BICM. 

\subsection{A comment on the case of non-integer $n_d$}
For memoryless transmission and likelihood function, the LLRs of all bits mapped to the partially-loaded position are the same. Rounding is used during optimization in order to ensure that the purely data-carrying positions have maximized MI. The effect of partially loaded positions in cases where $n_d$ is \textit{rounded down} is that some dummy bits will exhibit non-zero MI, but are discarded anyway. This leads to wasted MI, which cannot be exploited for data transmission. The effect in cases where $n_d$ is \textit{rounded up} is that some data bits will exhibit zero MI. From the decoder's perspective, these bits are equivalent to \textit{erasures}. In this paper, they are not treated differently than all other coded bits. However, for future work, erasure-aware coding techniques can be exploited to improve the performance further. 

We also mention that constellations optimized by always rounding down or up instead of to nearest integer have been the subject of our investigation as part of this work. However, we found that nearest integer provides the best trade-off between the two cases described above. This will be discussed further in \sec{exp_results}. 

\subsection{Time-sharing hybrid modulation aspects}
The drawback of the partially loaded bit can be mitigated by the adoption of a TH modulation. Here, for the $n_d-\lfloor n_d \rfloor$ fraction of the time, the constellation optimized for $\lceil n_d \rceil$ is used, and for the remaining $\lceil n_d \rceil - n_d$ fraction, the constellation optimized for $\lfloor n_d \rfloor$ is used. When TH is allowed, two constellations need to be optimized, and two different constellations need to be alternating at the demapper. This may lead to increased logic complexity at the receiver side, as well as increased complexity of potentially decision-directed digital signal processing (DSP) algorithms, for example phase noise recovery using a blind phase search. Nevertheless, TH is expected to perform better than MTOM due to the lack of erasures. The AIR in this case is given by
\begin{align}
 \label{eq:TH_AIR}
 & AIR = \notag \\
 & (n_d-\lfloor n_d \rfloor)\cdot \left(m-\lceil n_d \rceil + \sum_{i=1}^{m-\lceil n_d \rceil} \E_k \left[\log_2 p(\mbf{u}_k^i|y_k) \right]\right) + \notag \\ 
 & (\lceil n_d \rceil - n_d)\cdot \left(m-\lfloor n_d \rfloor + \sum_{i=1}^{m-\lfloor n_d \rfloor} \E_l \left[\log_2 p(\mbf{u}_l^i|y_l) \right]\right),
\end{align}
where the 1st and 2nd row of \eq{TH_AIR} correspond to the AIR for the time instants $k$ where $\lceil n_d \rceil$ is used and time instants $l$ where $\lfloor n_d \rfloor$ is used, respectively. Using automatic differentiation for the optimization of a TH scheme is directly possible by setting the OF to the AIR in \eq{TH_AIR}, and optimizing two constellations jointly. The two constellations are transmitted at different time slots, and batch symbols $k$ are connected to the cost function only through the constellation optimized for $\lceil n_d \rceil$ (and vice versa for symbols $l$). The GCS TH optimization is therefore equivalent to optimizing the two constellations independently and applying them during their respective required fractions of time. 

In conventional TH modulation, the modulation format \textit{size} also changes, but the requirement for 'dummy' bits is fully relaxed. This is also supported with the proposed scheme, and may be slightly beneficial due to the fewer number of optimization variables, leading to potentially lower probability of local optimum. 

\subsection{A comment on similarities with autoencoders}
The method of AD has recently been applied for optimization of autoencoder (AE)-based constellation shaping \cite{Hoydis, RodeRobust, OgnjenRobust, Jones2019, Karanov2019}. The optimization variables in those cases are the weights of the encoder and decoder NNs. As demonstrated above, the decoder part of the AE is not necessary when the channel is Gaussian and the variance is known. Furthermore, an encoder is also not necessary because the transmitter does not compress the binary data - the entropy of the signal defined by $\X$ is the same as the sum of the entropies of the bits. Instead, as described, the fixed LUT-based labeling can be used without any penalty. The decoder part of the AE structure is of interest when the channel is nonlinear and gives rise to an unknown and/or difficult to approximate likelihood function (e.g. due to very high complexity). The encoder is of interest when compression is employed at the transmitter, or when the cardinality prohibits LUT-based mapping (e.g. in the case of massive multiple-input multiple-output communications). Neither is the case for the majority of communication systems where the data compression is independent of the transmission and the transmitted symbols are typically corrupted by Gaussian noise (after DSP processing).

\begin{figure*}[!t]
\centering
\subfigure{
 \includegraphics[trim=0 0 0 0cm, width=0.23\linewidth]{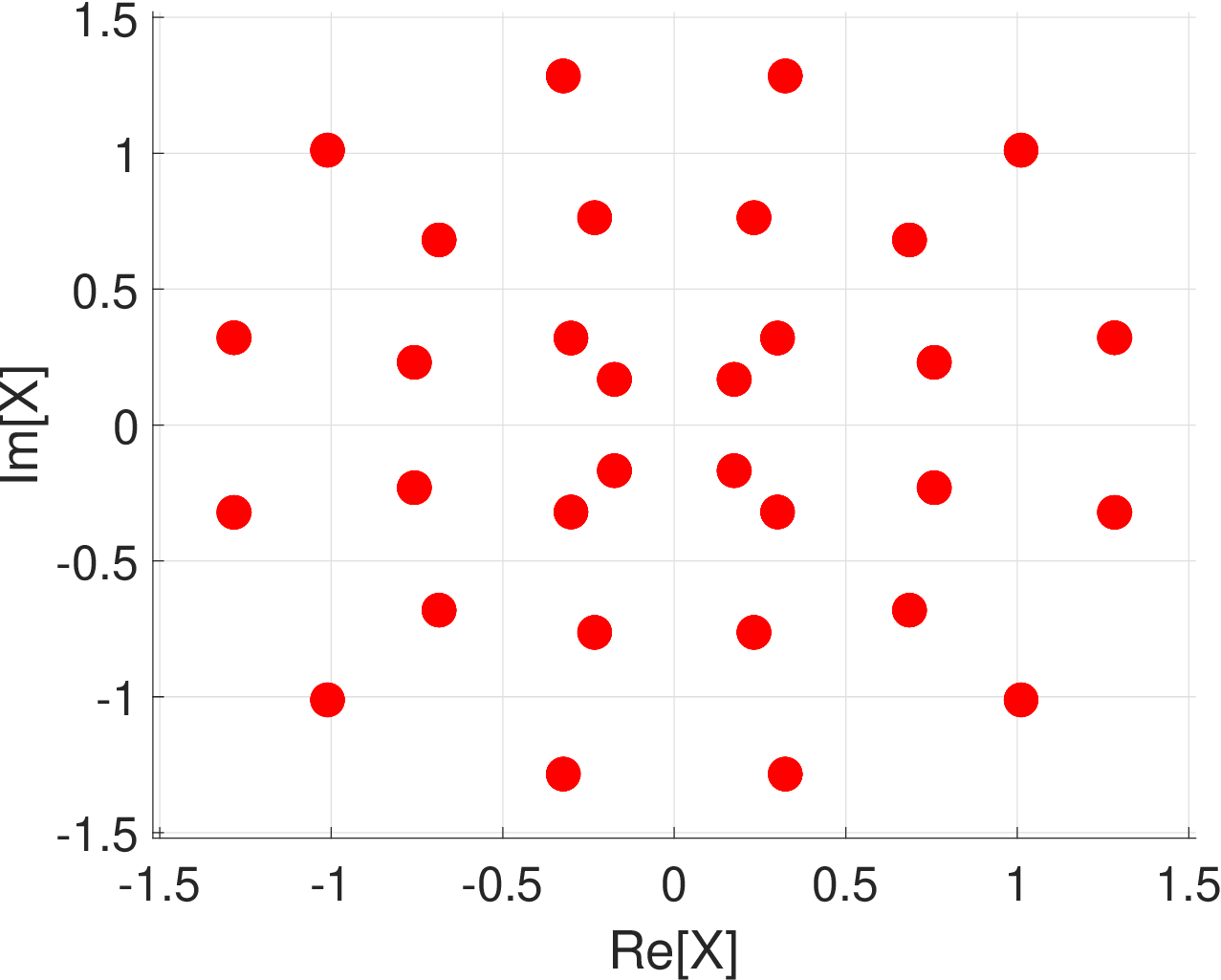}}
\subfigure{
 \includegraphics[trim=0 0 0 0cm, width=0.23\linewidth]{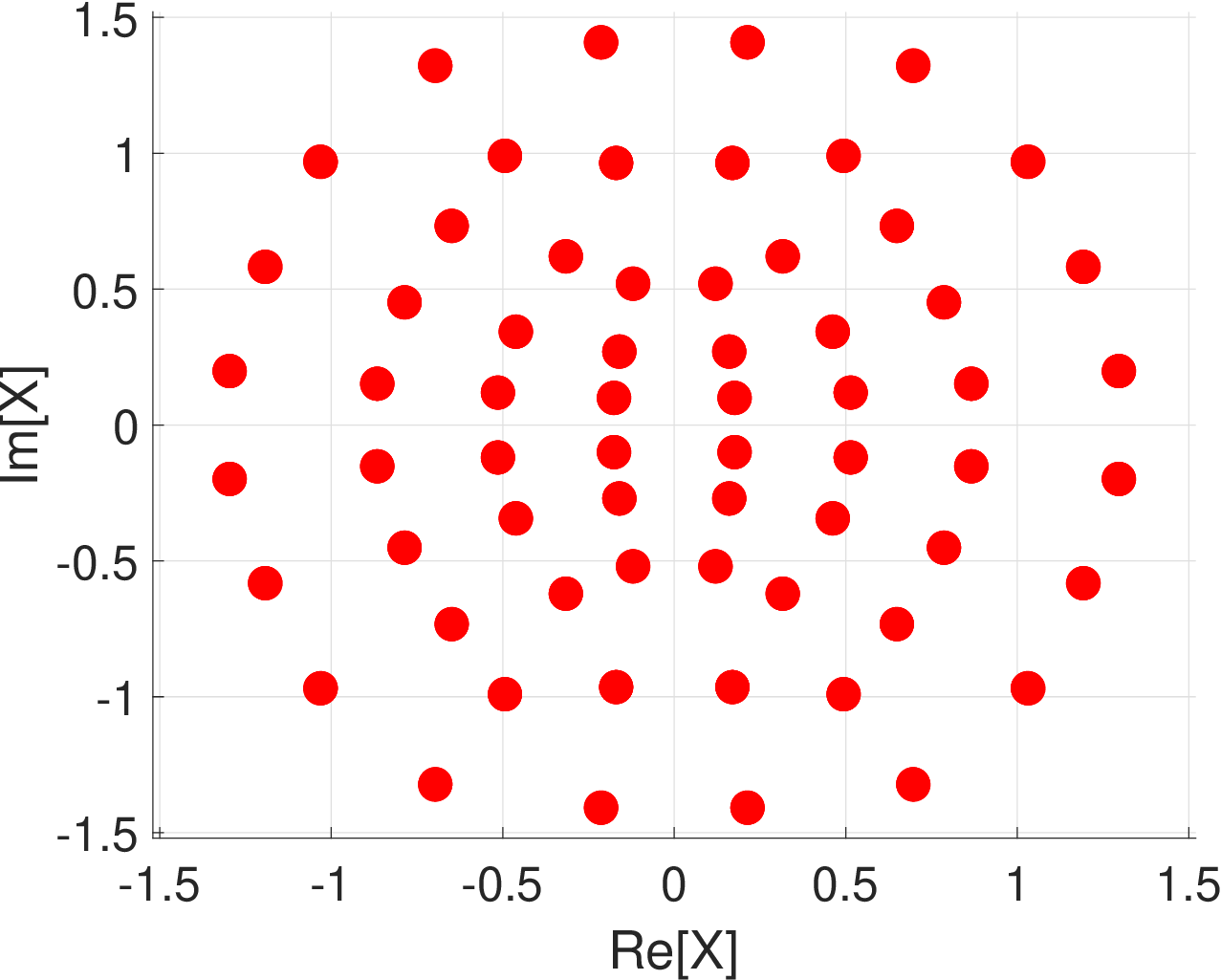}}
 \subfigure{
 \includegraphics[trim=0 0 0 0cm, width=0.23\linewidth]{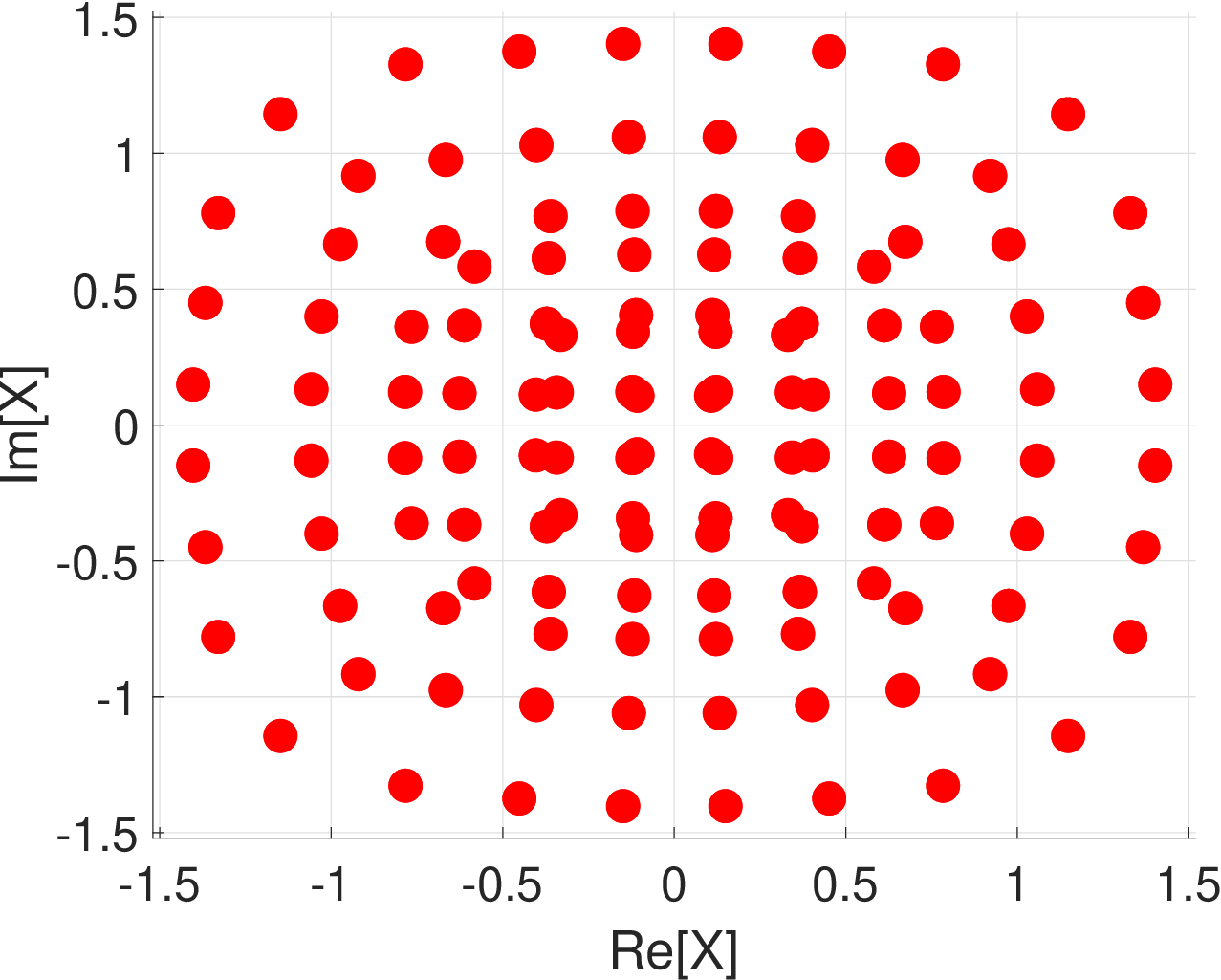}}
 \subfigure{
 \includegraphics[trim=0 0 0 0cm, width=0.23\linewidth]{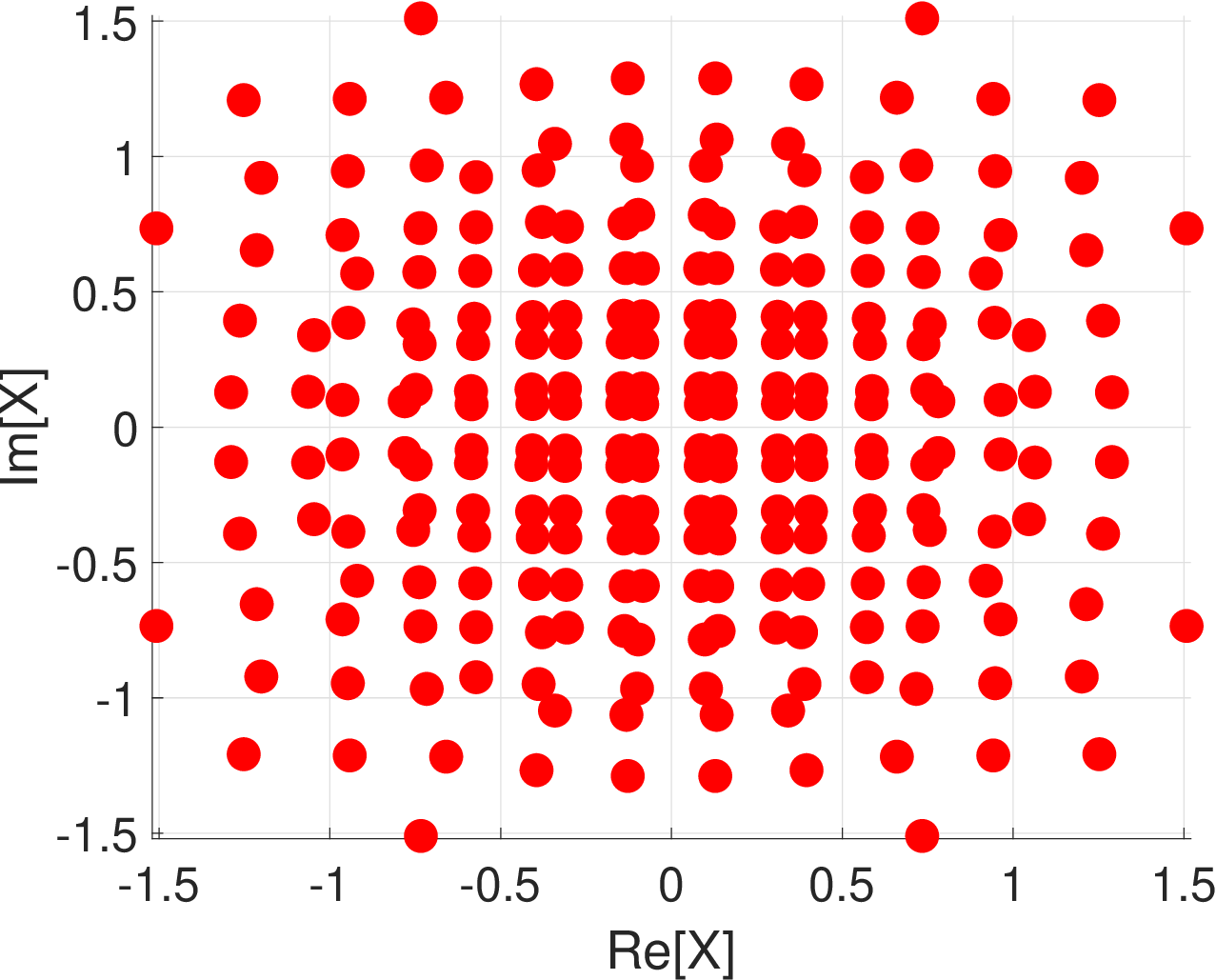}}
 \caption{Optimized constellations for MTOM with $n_d=3,2,1,0$ and SNR=13,15,17,19 dB from left to right, respectively.}
 \label{fig:const_AWGN}
\end{figure*}

\begin{figure*}[!t]
\centering
\subfigure{
 \includegraphics[trim=0 0 0 0cm, width=0.31\linewidth]{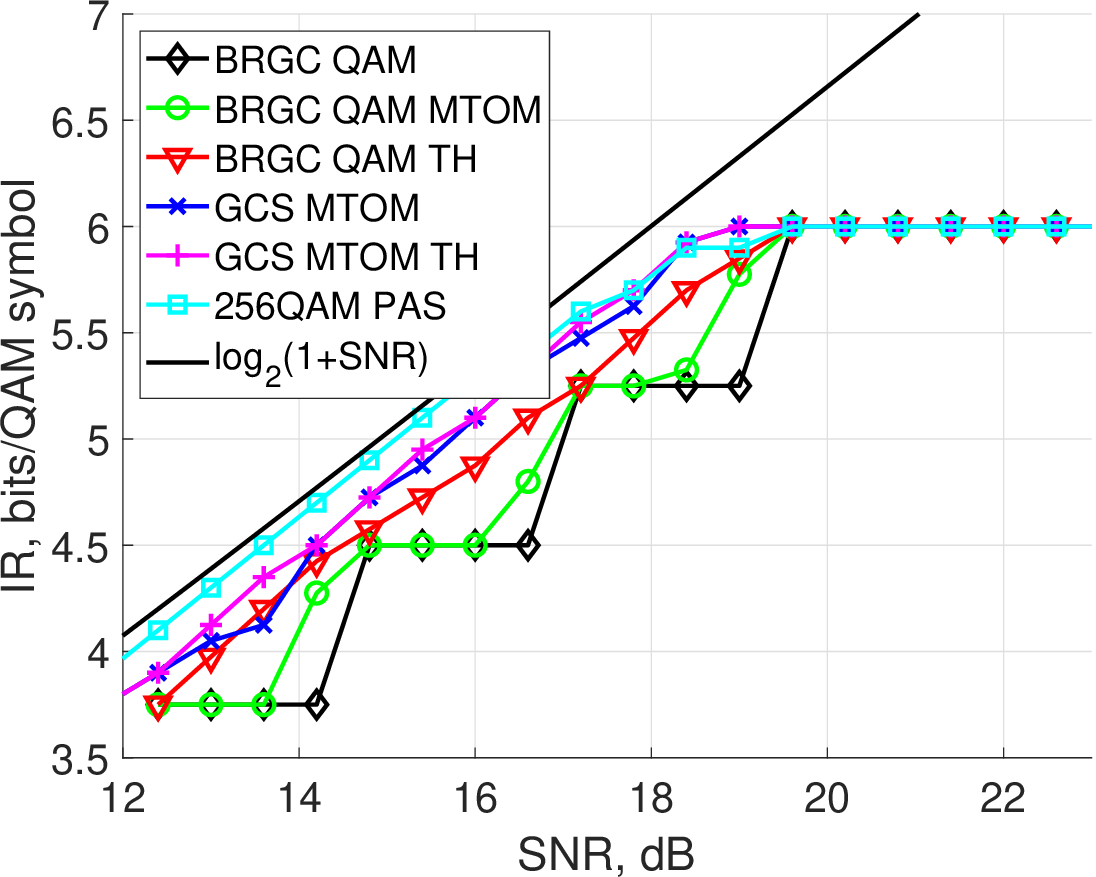}}
 \subfigure{
 \includegraphics[trim=0 0 0 0cm, width=0.31\linewidth]{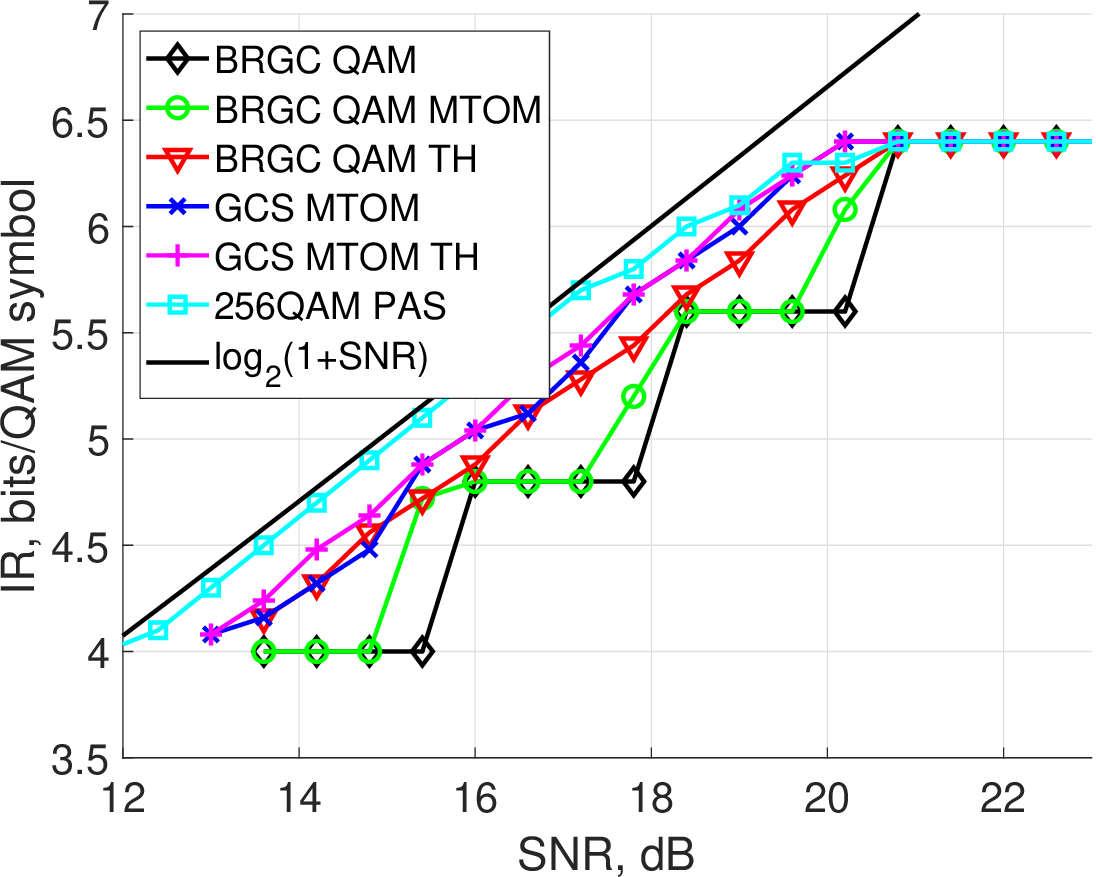}}
 \subfigure{
 \includegraphics[trim=0 0 0 0cm, width=0.31\linewidth]{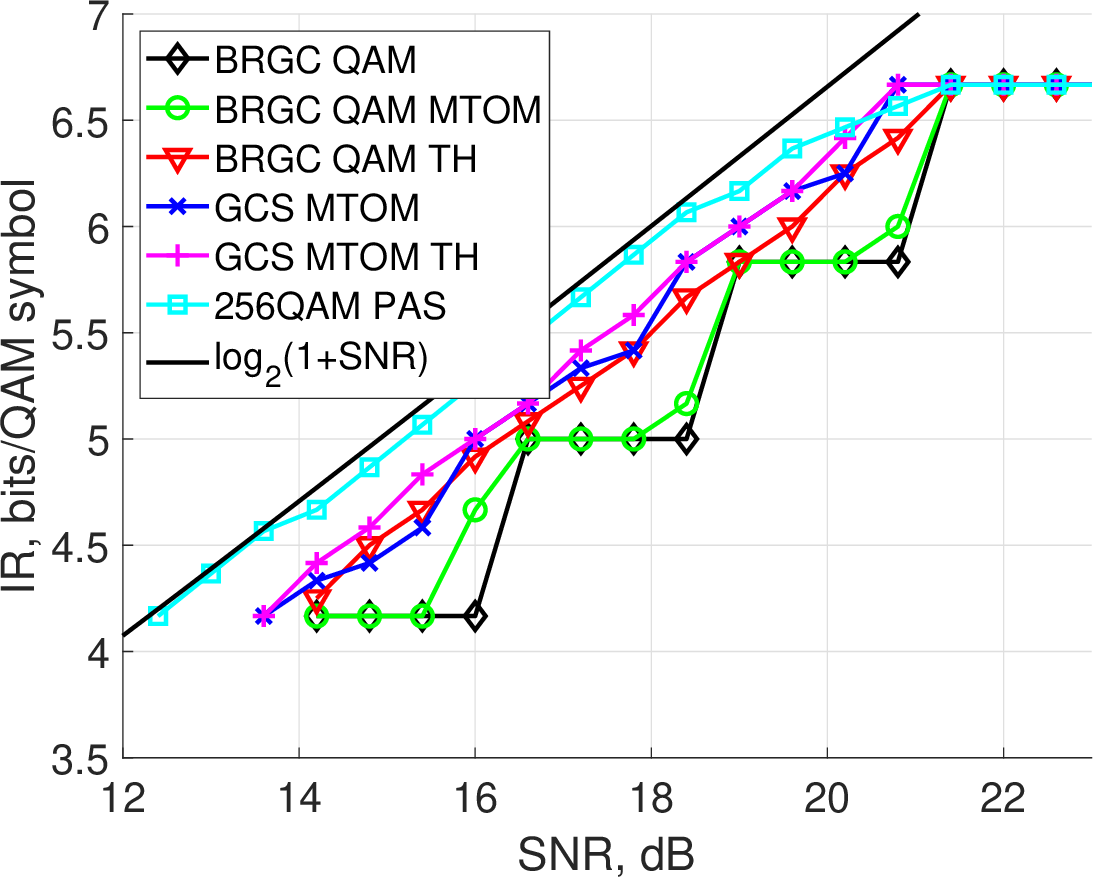}}
 \\
 \subfigure{
 \includegraphics[trim=0 0 0 0cm, width=0.31\linewidth]{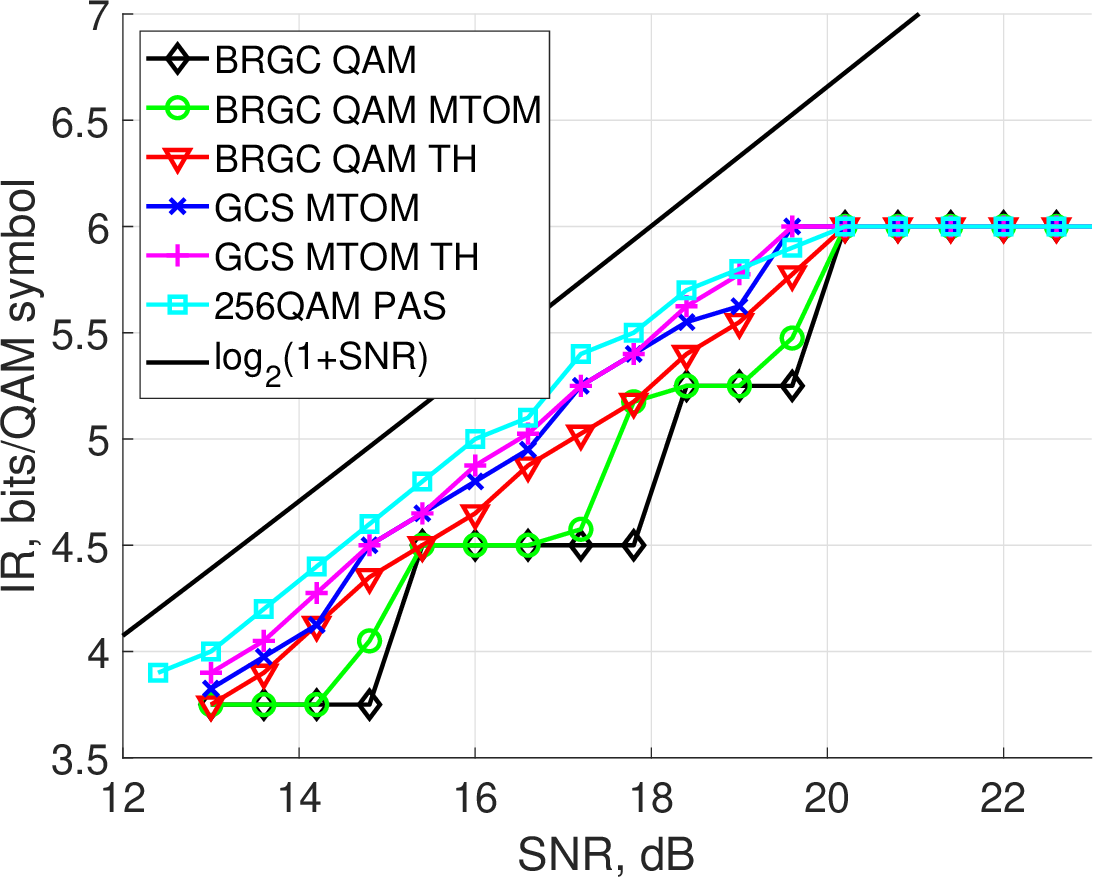}}
 \subfigure{
 \includegraphics[trim=0 0 0 0cm, width=0.31\linewidth]{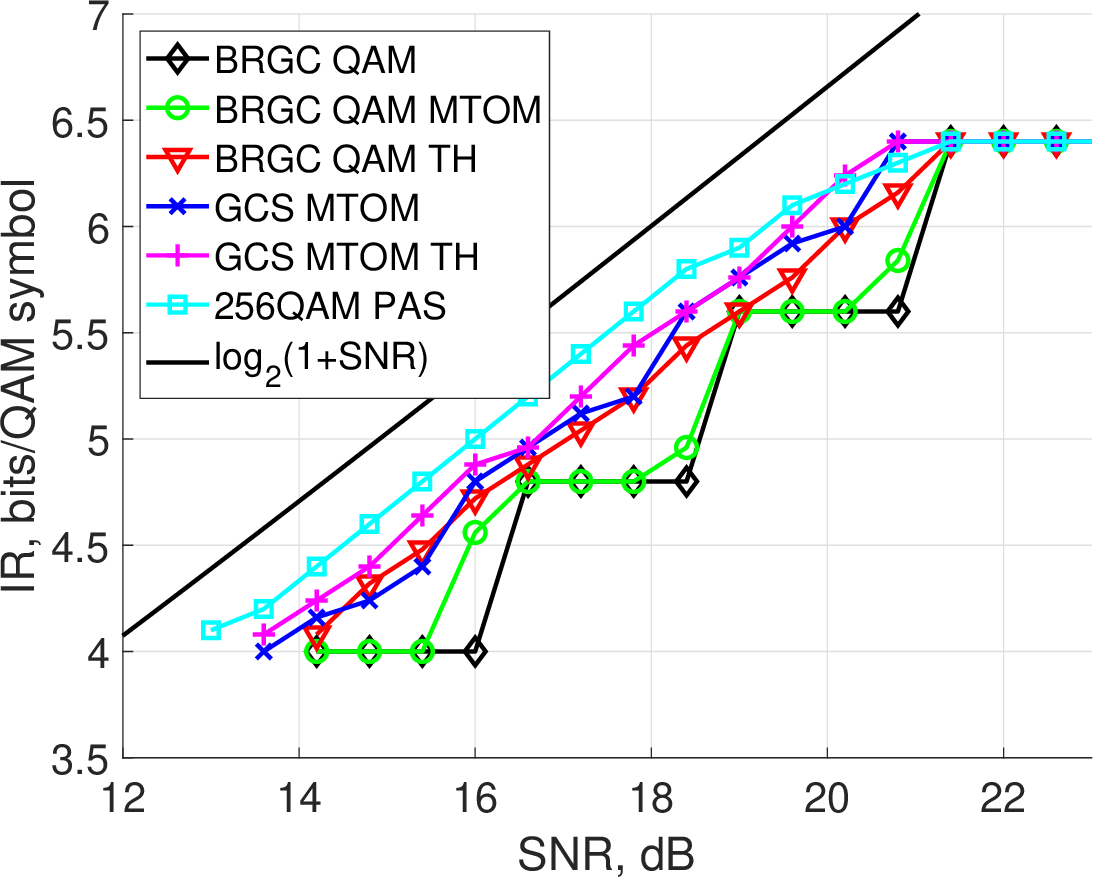}}
 \subfigure{
 \includegraphics[trim=0 0 0 0cm, width=0.31\linewidth]{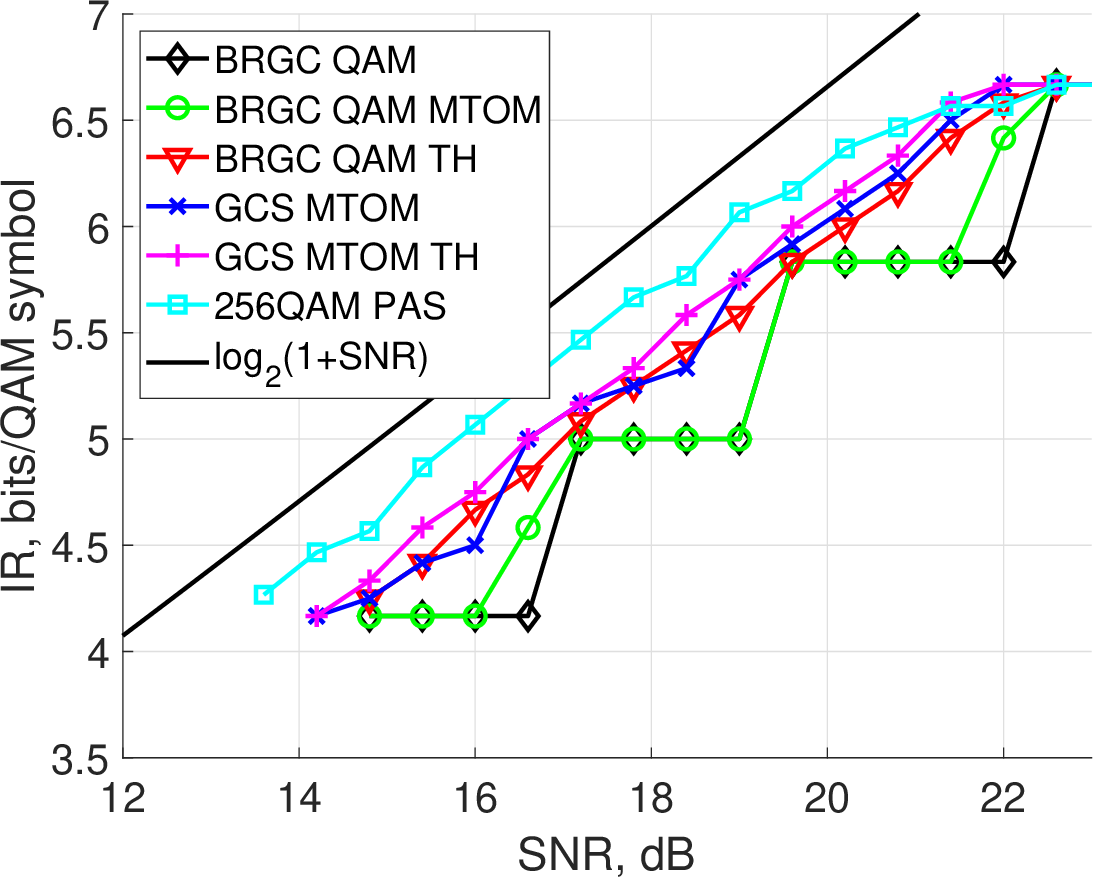}}
 \caption{Simulation results for minimum SNR required to achieve the target rate. \textbf{Top row:} assuming ideal FEC based on minimum SNR where AIR $>$ IR; \textbf{Bottom row:} with LDPC FEC, where $BER < 10^{-5}$ at the target IR. Results for 33\%, 25\% and 20\% FEC overhead in the first, second and third column, respectively.}
 \label{fig:sim_results_AWGN}
\end{figure*}

\section{Results for AWGN channel}
\label{sec:AWGN_results}
First, the simplest case of AWGN channel is studied, where NLIN and quantization noises are excluded from the model. For optimization, the stochastic gradient descent algorithm is employed with the Adam algorithm. A total of 100.000 symbols are used for training, separated into batches of 500 symbols. The learning rate was set to $10^{-3}$ and an L2 regularization was employed with a weight decay of $10^{-5}$. These numbers were optimized by a grid search. The target IR is studied for the MTOM scheme with $n_d \in \left[0, \min\{3, m-5\}\right]$ with a step of 0.05. The case of $n_d=0$ implies conventional BICM. Optimization using AD is performed for 256QAM. The conventional 64QAM and 256QAM are also studied with BRGC, as well as 32QAM and 128QAM with a minimum Gray penalty labeling scheme \cite{crossQAM} (where we slightly abuse the terminology and still referred to the labeling of 32QAM and 128QAM as BRGC). 

The total number of dummy bits for $N=64800$ is $N_D \in \left[0; 38880 \right]$. The FEC rates studied are $R=3/4$, $R=4/5$ and $R=5/6$ (i.e. overheads of 33\%, 25\% and 20\%), all from the DVBS-2 family \cite{DVBS2}. A TH modulation is also evaluated with and without GCS. The PAS scheme applying a constant composition distribution matcher (CCDM) \cite{Schulte} is also studied. The rate of PAS is swept by varying the entropy $\H(X)$ of the MB PMF in the range $\left[5; 8\right]$ with a step of 0.1 for 256QAM. The net data rate in this case is given by $IR=\H(X) - m\cdot(1-R)$. The MB PMF is a near-optimal family of distributions for the AWGN channels \cite{Frank}, and the CCDM of length $>1000$ implies minimal rate loss \cite{Schulte}. It was verified that at the target rates, 256QAM is superior to 64QAM in an AWGN setting. 

The constellations optimized with $n_d=3,2,1,0$ to be used at $SNR=13,15,17,19$ dB, respectively, are given in \fig{const_AWGN}. All constellations contain 256 points, and are normalized to unit power, leading to peak-to-average power ratios (PAPRs) of 2.05, 2.24, 2.63 and 3.03, respectively. As an example, at $n_d=3$, constellation points are grouped into groups of $2^{m-n_d}=8$ with nearly identical coordinates. The points within each group are determined by the three LRB and cannot be distinguished. However, they also do not carry information and are discarded at the receiver. Similarly, for $n_d=2$ and $n_d=1$ the points are grouped into groups of 4 and 2, respectively. Further examples are given in the rest of the paper. 

In \fig{sim_results_AWGN}, the minimum SNR required to achieve error-free performance is reported. On the top row, the idealized case is shown, where a rate is considered achieved if the AIR is above it. On the bottom row, the practical FEC-based case is shown, where an error-free performance is assumed when the BER after LDPC decoding is $<10^{-5}$. The performance is based on a simulation of 1000 LDPC codewords in each case. At every SNR, the maximum IR which is achieved without errors is reported as per the above-stated definitions. The SNR is swept with a step of $0.6$ dB. 

As example, the case of $R=3/4$ is discussed in detail. The conventional BRGC QAM achieves AIR-based error-free performance for $SNR \ge 12.4, 14.8, 17.2 \text{ and } 19.6$ dB. Here and in the other sub-figures of \fig{sim_results_AWGN}, the four 'steps' in performance correspond to the rate with 32QAM, 64QAM, 128QAM and 256QAM, respectively. The penalty of the FEC is estimated as the differences between the top and bottom row of \fig{sim_results_AWGN} and is $\approx 1$ dB in all studied cases. The MTOM without shaping allows for slight improvement w.r.t. conventional BRGC QAM. The TH without shaping achieves a smooth performance as a function of the SNR. The proposed scheme also achieves smooth rate adaptivity and a shaping gain w.r.t. the TH without shaping of up to 0.7 dB, and decreasing at low SNR. Applying a TH to the GCS allows for further improvement in the performance due to the above-mentioned lack of erased/punctured bits. The proposed scheme appears most effective at the relatively low $R$, where the penalty of asymmetric labeling appears to be smallest. As expected, and also suggested in a previous study comparing PAS to GCS in an AWGN setting \cite{Fabian}, PAS is more effective when the constellation size is finite due to its near-capacity PMF and the ability to Gray-code. Additional up to 1 dB of gain can be achieved using PAS w.r.t. the TH with GCS depending on the rate. A minor drawback of PAS is that at the high net data rates, it is ultimately limited by $R$ and has to apply a near-uniform PMF, thereby foregoing of its shaping gain (unless a higher order modulation is selected). The proposed scheme achieves GCS gain also at the high target rates. Additional discussion is provided in \sec{discussion}. The rest of the paper focuses on the proposed MTOM scheme, and the TH versions are therefore excluded from further studies.

\section{Results for the ideal fiber channel model}
\label{sec:ideal_results}
In this section, the AIRs are studied for the fber channel used for optimization. The purpose of this study is to obtain the potential of the MTOM scheme in an idealized scenario when not subjected to practical FECs and digital signal processing (DSP) techniques. To that end, the NLIN channel model \cite{NLIN}, which is similar to the EGN model is studied with standard single mode fiber (SSMF) parameters of loss $\alpha=0.2$ dB/km, dispersion $D=17$ ps/nm/km and $\gamma=1.3$ 1/W/km. Five channel WDM system is considered with 32 GBd per channel on a 50 GHz grid and a square-root cosine roll-off pulse shape with a roll-off factor of 0.01. Additionally, 8-bit quantization noise is applied after the transmitter and before the recevier as described above. An arbitrarily chosen SNR penalty of $SNR_{TRX}=35$ dB is also applied. A multi-span system is simulated with 100 km per span and an erbium doped fiber amplifier (EDFA)-based lumped amplification with an EDFA NF of 5 dB. In the case of conventional BRGC QAM and PCS, the optimal launch power is obtained by a launch power sweep for each distance with a step of 0.1 dBm. In this case, PCS is realized by using an uncoded, memoryless MB PMF source, whose entropy is optimized similar to the AWGN case, and it was similarly verified that 256QAM is superior to 64QAM. In the case of AD-based optimization, the launch power is added to the optimization tensor. The performance in all cases is based on the AIR from Eq.~\eq{AIR} averaged over 5 blocks of 100.000 symbols each. Each constellation for each target rate is evaluated for transmission distances up to 30 spans with a step of 1 span.    

\begin{figure}[!t]
\centering
 \includegraphics[trim=0 0 0 0cm, width=1.0\linewidth]{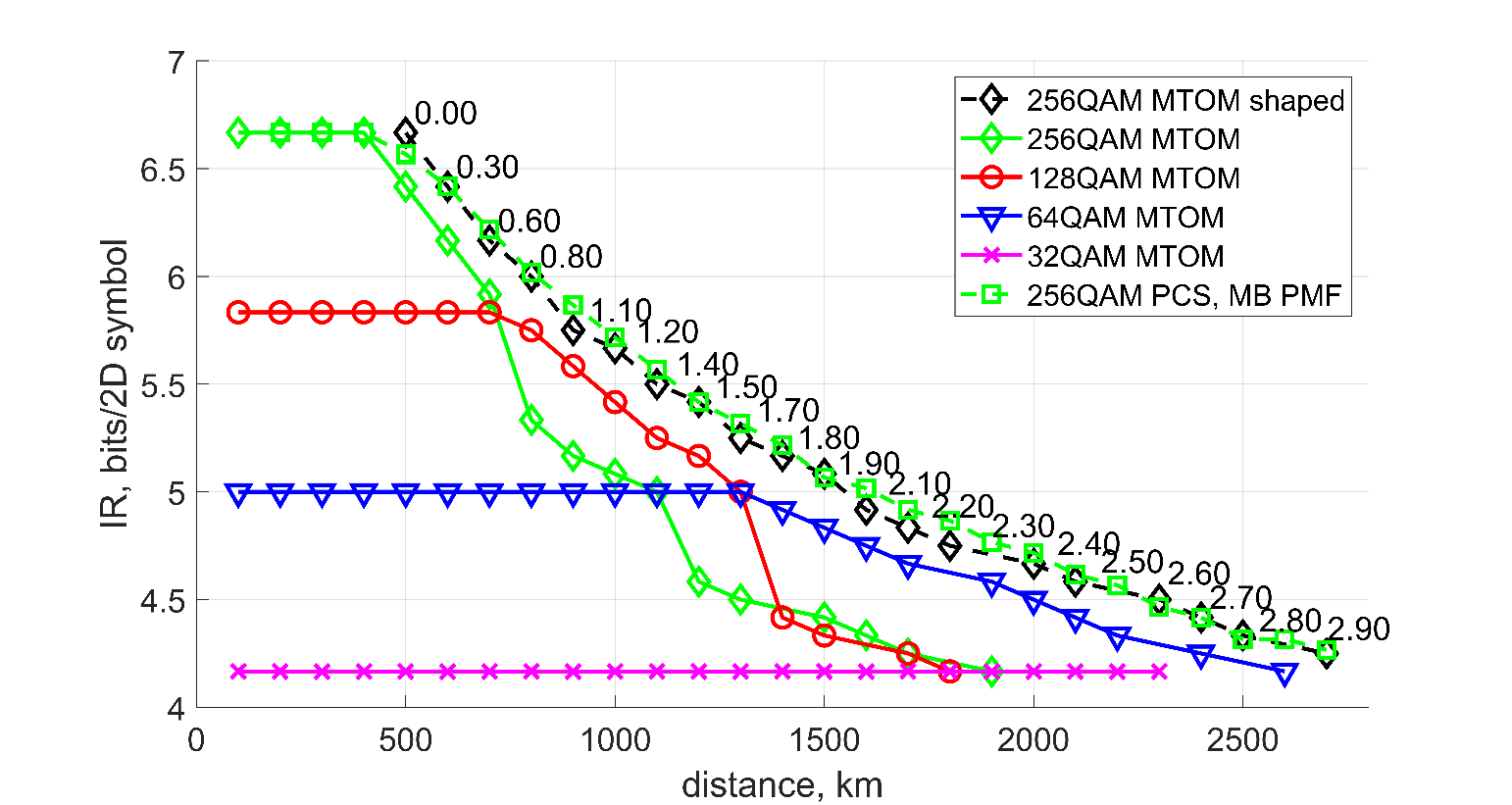}
 \caption{Achievable information rates for the channel model simulated in \sec{ideal_results}. The maximum distance at which the AIR is higher than the target IR is shown. For the shaped system, the corresponding average number of dummy bits per symbol $n_d$ is annotated.}
 \label{fig:results_NLIN}
\end{figure}

\begin{figure}[!t]
\centering
 \includegraphics[trim=0 0 0 0cm, width=1.0\linewidth]{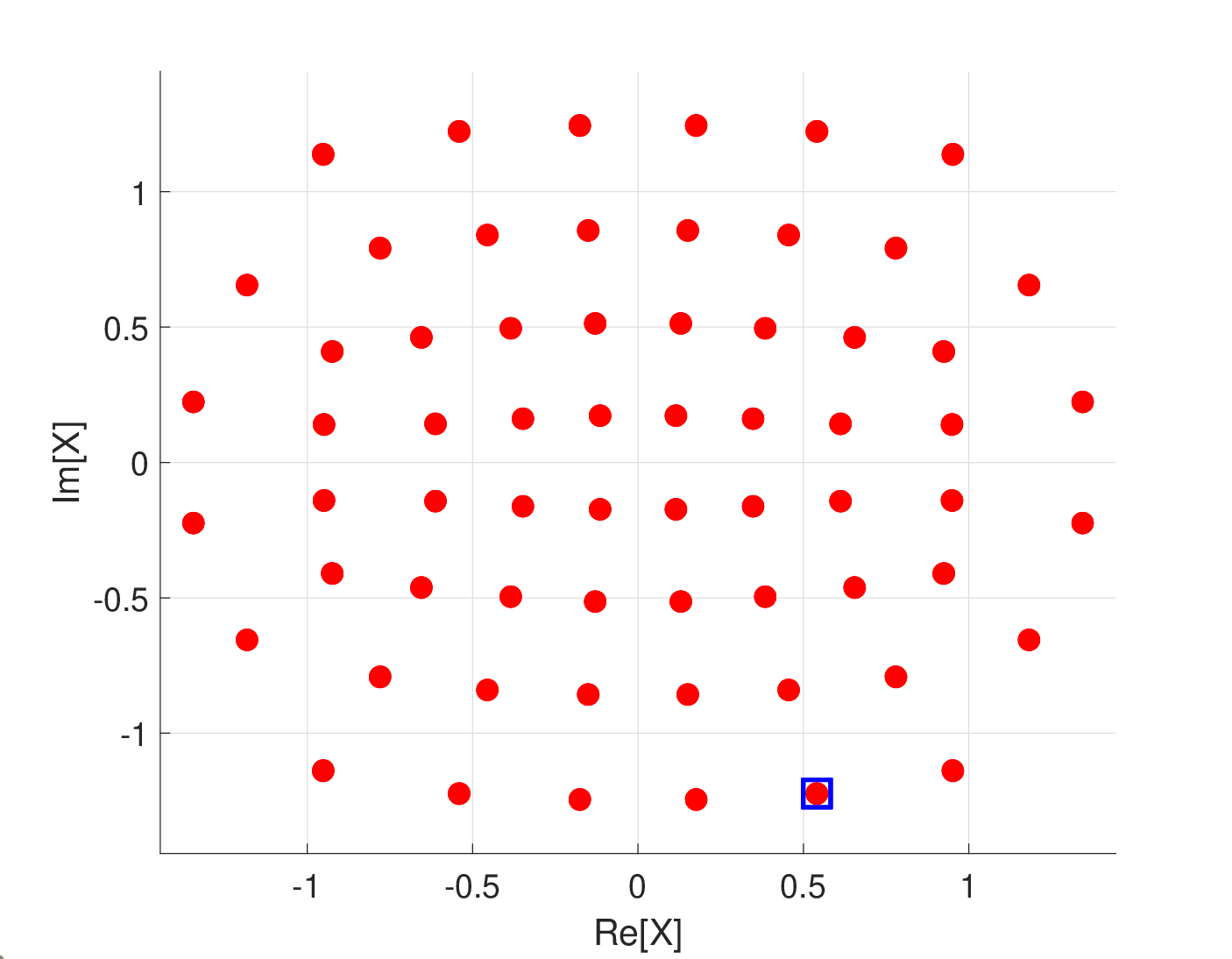}
 \caption{Example optimized constellation for 1500 km with $n_d=2$ for the channel model described in \sec{ideal_results}. The highlighted point is zoomed-in around in \fig{example_constellation_NLIN_zoom}}
 \label{fig:example_constellation_NLIN}
\end{figure}

\begin{figure}[!t]
\centering
 \includegraphics[trim=0 0 0 0cm, width=1.0\linewidth]{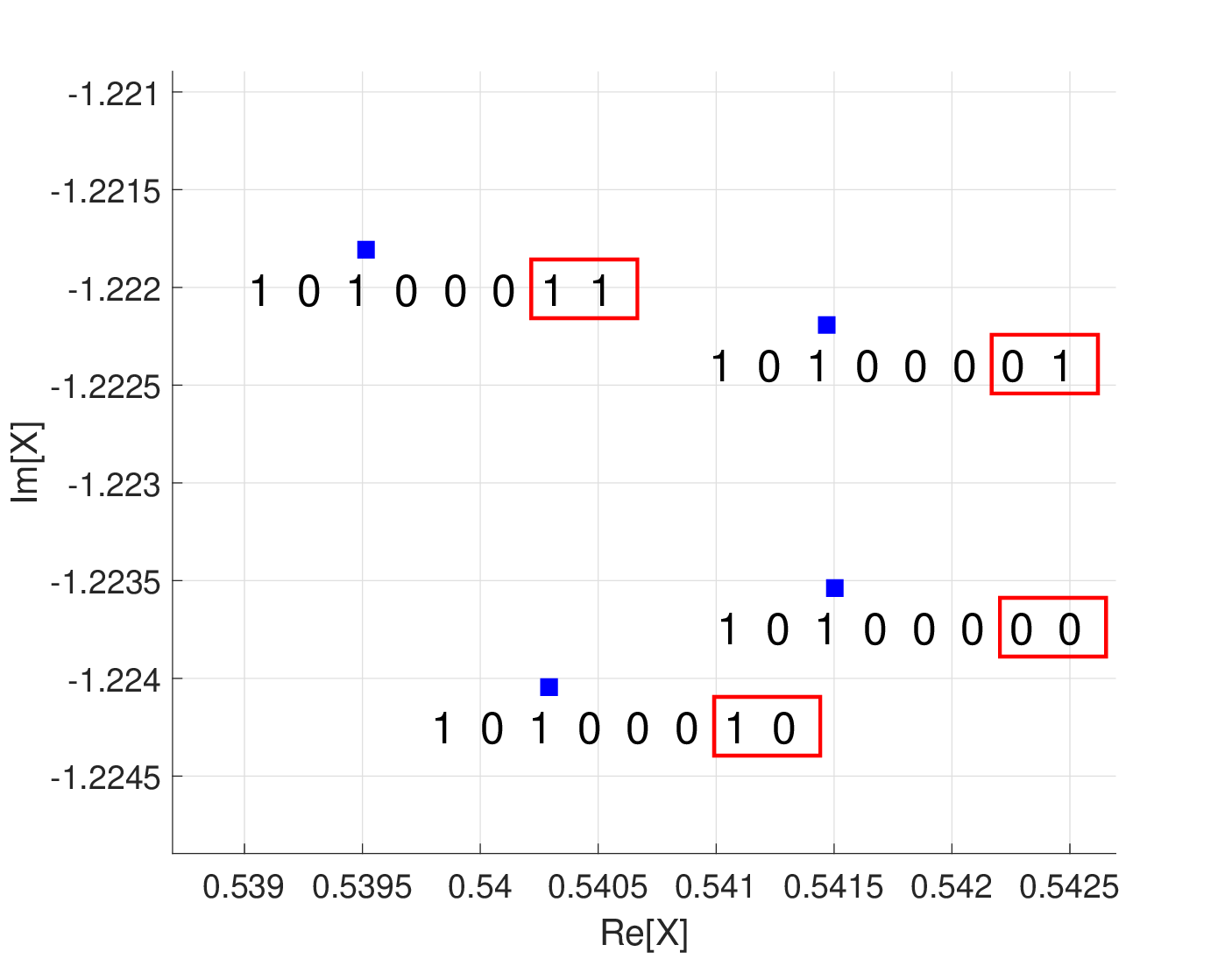}
 \caption{A zoom-in around the point identified in \fig{example_constellation_NLIN}, revealing the four points in close proximity to each other, carrying identical leading six bits in their labels.}
 \label{fig:example_constellation_NLIN_zoom}
\end{figure}

In \fig{results_NLIN}, the maximum distance at which the AIR is higher than the target rate is shown for each studied test case. This would correspond to the maximum distance at which the IR is achieved without errors with a capacity-achieving FEC. An FEC rate of $R=5/6$ is assumed. An example of the constellation optimized for 1500 km with $n_d=2$ is shown in \fig{example_constellation_NLIN}. The PAPR in this case is 2.20, corresponding to approx. 0.1 dB reduction w.r.t. the PAPR of 2.24 mentioned in the AWGN channel for the same $n_d$. A zoom-in around one of the points is shown in \fig{example_constellation_NLIN_zoom}, revealing the four points residing at a very close proximity to each other, as well as their labels, which are identical in the first six positions. The MI per bit for this case is given in Table~\ref{tbl:MI_per_bit}, together with the MI per bit for BRGC 256QAM, 128QAM and 64QAM, as well as a 256QAM PAS with an entropy $\H(X)=6.4$ bits/2D symbol and a target rate of $5.06$ bits/2D symbol. Even though the MI on the `dummy' positions of the shaped QAM is 0, the MI on the other positions is enhanced, leading to an increased total AIR of 5.06 bits/2D symbol, which is just over the target $IR=R\cdot(m-n_d)=5$. On the nonlinear channel, PCS achieves similar performance to the proposed GCS, where the MB PMF appears to be penalized due to the high PAPR and the corresponding NLIN. 

\begin{table}
\scriptsize
 \caption{\scriptsize{MI per bit for BRGC QAM without shaping, GCS QAM with $n_d=2$ for the constellation from \fig{example_constellation_NLIN}, and 256QAM PAS with an MB PMF with an entropy of 6.40 bits/2D symbol simulated using the channel model described in \sec{ideal_results}}.}
 \label{tbl:MI_per_bit}
 \centering
 \begin{tabular}{c||c|c|c|c|c}
  bit $i$& 64QAM & 128QAM & 256QAM & 256QAM GCS & 256QAM PAS \\
  \hline
  1 & 0.92 & 0.89 & 0.88 & 0.91 & 0.75 \\ 
  2 & 0.84 & 0.73 &  0.77 & 0.92 & 0.68 \\
  3 & 0.68 & 0.79 &  0.55 & 0.78 & 0.64 \\
  4 & 0.92 & 0.44 &  0.20 & 0.86 & 0.48 \\
  5 & 0.84 & 0.89 &  0.88 & 0.72 & 0.75 \\
  6 & 0.68 & 0.66 &  0.77 & 0.85 & 0.68 \\
  7 & n/a & 0.44 &  0.55 & 0.00 & 0.64 \\
  8 & n/a & n/a  & 0.20 & 0.00 & 0.48 \\
 \hline
 \hline
 Total & 4.89 & 4.85 & 4.83 & 5.06 & 5.12
 \end{tabular}
\end{table}

A summary of the PAPR and constellation higher order moments is given in Table~\ref{tbl:moments} for the conventinoal 256QAM, the 256QAM with $n_d=2$ shaped for the AWGN channel, the 256QAM with $n_d=2$ shaped for the NLIN channel, and the above-mentioned 256QAM with a MB PMF of entropy 6.4 bits/2D symbol. The constellation optimized for NLIN channel exhibits lower moments than the one optimized for the AWGN channel, but slightly higher than conventional QAM, indicating that tolerance to the AWGN noise remains important alongside minimizing NLIN for maximizing the objective function. 

Unshaped constellations employing MTOM are competitive for cases of $n_d<1$, which can be seen in \fig{results_NLIN} with the sharp degradation of the performance of 256QAM and 128QAM after 800 km and 1300 km, respectively (rates of $R\cdot (m-1)$). After these thresholds are reached, the fact that points with identical labels on the data carrying positions are not merged begins to penalize the MI performance on those positions. When the constellation is shaped, points are merged which allows to exploit the zero-MI of dummy bits to maximize the MI of data carrying bits. Some more constellation examples will be seen in \sec{exp_results}. The shaping gain is $\approx 2$ spans or $\approx 0.2$ bits/2D symbol, and is maintained throughout the studied target IRs and distances.

\begin{table}
\scriptsize
 \caption{\scriptsize{Statistics for selected 256QAM constellations. All constellations normalized to unit power}.}
 \label{tbl:moments}
 \centering
 \begin{tabular}{c||c|c|c|}
  Coonstellation & $\max_i\left[|\X^i|^2\right]$ & $\E_i\left[|\X^i|^4\right]$ & $\E_i\left[|\X^i|^6\right]$ \\
  \hline
  Conventional & 2.65 & 1.39 & 2.29  \\ 
  GCS, $n_d=2$, AWGN & 2.24 & 1.52 & 2.58  \\
  GCS, $n_d=2$, NLIN & 2.20 & 1.44 & 2.38  \\
  MB PMF, $\H(X)=6.4$ & 11.29 & 1.98 & 5.74  \\
 \hline
 \hline
 \end{tabular}
\end{table}

\section{Experimental setup}
\label{sec:setup}
\begin{figure*}[!t]
\centering
 \includegraphics[trim=0 0 0 0cm, width=1.0\linewidth]{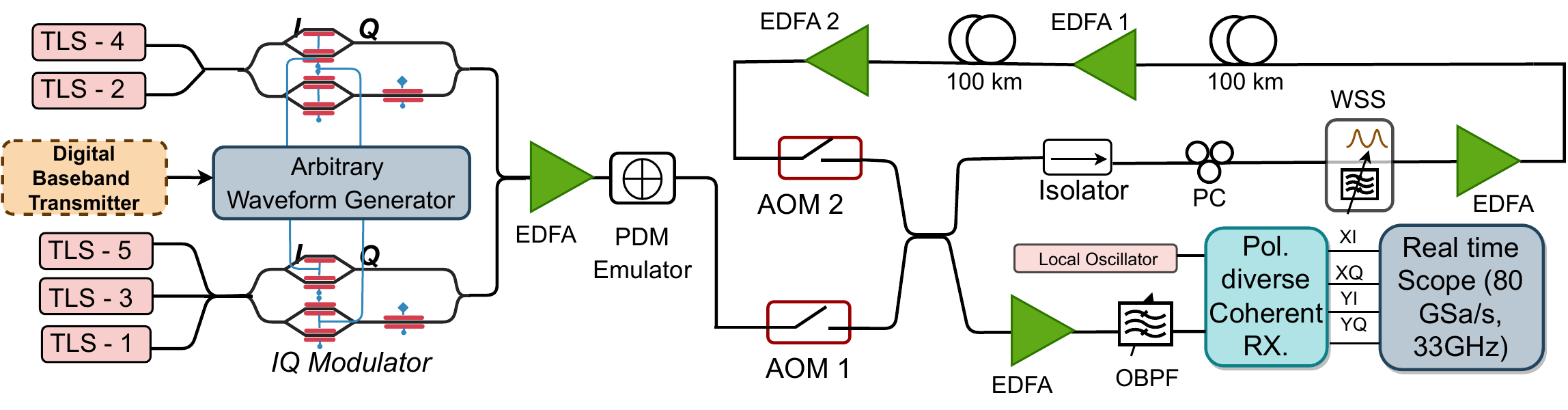}
 \caption{Experimental setup for transmission.}
 \label{fig:setup}
\end{figure*}

The experimental setup is given in \fig{setup}. The QAM symbols with a baudrate of 32 GBd are pulse shaped by a square-root cosine roll-off pulse with a roll-off factor of 0.01, and sent to an arbitrary waveform generator (AWG) with a 64 GSa/s rate. Two independent signals drive two in-phase in-quadrature (IQ) modulators for the odd channels and the even channels of the 5-channel WDM link, respectively. The five transmitter laser sources (TLSs) are external cavity lasers with a linewidth of 10 kHz and are placed on a 50 GHz grid. The central channel is at 192.5 THz. Polarization division multiplexing emulation is applied with the time decorrelation method by $>1880$ symbols. The multi-span link is realized by a recirculating loop. The loop consists of two spans of 100 km TeraWave® SCUBA fiber each. The data sheet parameters of the fiber are loss $\alpha=0.155$ dB/km, nonlinear coefficient $\gamma=0.715$ 1/W/km and a dispersion coefficient $D=22$ ps/nm/km. The span losses are compensated with two EDFAs. A wavelength selective switch (WSS) is integrated in the loop to suppress out-of-band noise accumulating during recirculations, as well as to flatten the gain spectrum at each loop turn. An additional EDFA is used to compensate the losses of the two acousto-optic modulators (AOMs) used for controlling the desired number of recirculations. A polarization controller (PC) is used to reduce the effect of any polarization dependent gain accumulated in the loop. The launch powers of $\{-0.9, 0.1,  0.5, 1.0,  1.6,  2.1,  2.6\}$ dBm per channel and distances between 200 and 3000 km with a step of 200 km (i.e. one loop turn) are evaluated.

\begin{figure}[!t]
\centering
 \includegraphics[trim=0 0 0 0cm, width=1.0\linewidth]{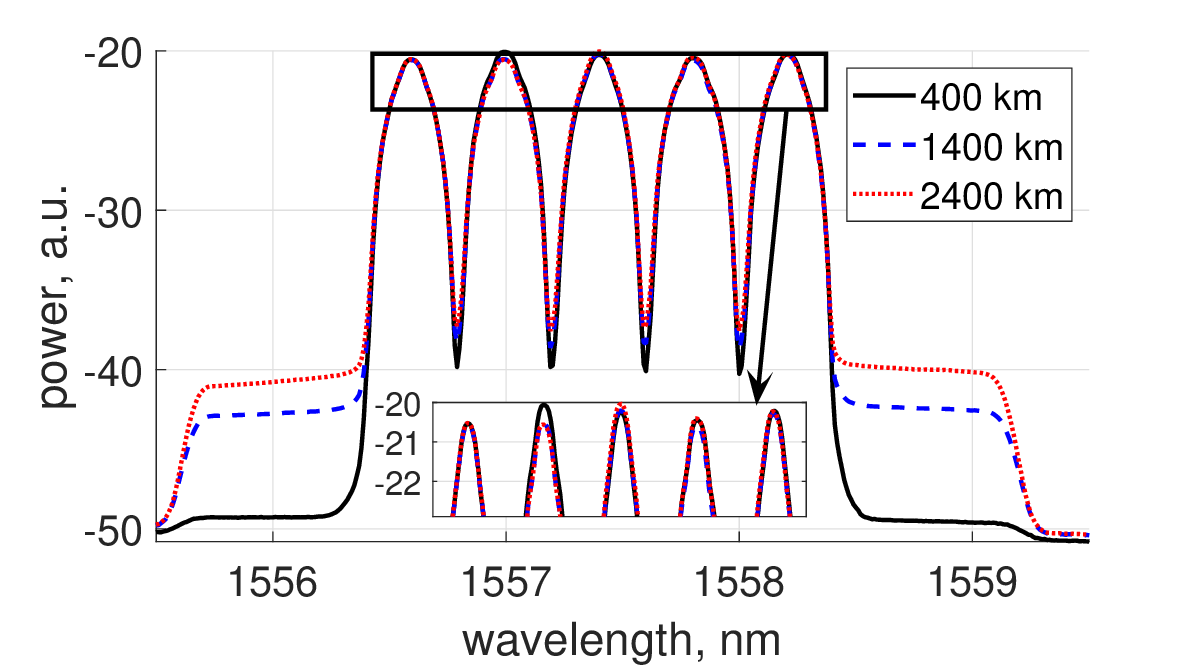}
 \caption{Example received spectra after transmission.}
 \label{fig:spectra}
\end{figure}

The received spectra are given in \fig{spectra} for the examples of 400 km, 1400 km and 2400 km, at the launch power of 0.5 dBm per channel. As can be seen at the inset, the WSS is tuned to achieve flatness in all cases to within 0.5 dB. 

After transmission, the WDM signal is passed through an optical band-pass filter (OBPF) to extract the central channel, which is then detected using a coherent receiver and an 80 GSa/s digital storage oscilloscope. The baseband signals are then processed offline. The received traces are first chromatic dispersion compensated in the frequency domain. Then, downsampling to two samples per symbol is performed using the Gardner synchronizer and a linear interpolation. A combination of training sequence and pilot-based carrier frequency offset estimation and compensation using the FFT method from \cite{Barbosa} are performed. The training sequence is used for coarse frequency estimation, and the pilots are used to compensate for frequency offset variations for the duration of the trace. Pilot aided radial decision based equalization is performed, where the equalizer is initialized based on a data-aided equalization for the training sequence duration. Then, downsampling to one sample per symbol is performed. Pilot aided carrier phase offset estimation is performed using the Tikhonov distribution based algorithm \cite{YankovJLT_PhaseNoise}, which assumes a Gaussian noise with a variance estimated from the training data at the beginning of each detected trace. The phase noise tracking algorithm directly produces symbol likelihoods. The symbol likelihoods are finally marginalized to obtain the LLRs, which are then fed to the FEC decoder.

The training sequence length is 2000 symbols, and the pilot rate is 1/32 similar to the ZR+ standard \cite{oFEC}.

\subsection{Parameter fit of the NLIN model}
In order to match the optimization model to the experimental test bed, NLIN parameter fit is performed for the parameters $SNR_{TRX}$, span EDFA NF, $\alpha$ and $\gamma$. The datasheet NF of the span EDFAs is $\approx 5$ dB. However, in order to avoid modeling the extra EDFA used for switch loss compensation, its effect will be included into the NF of the span amplifiers. It is therefore expected that the NLIN model fit will use an EDFA of a higher NF than the span EDFAs used in the experiment.

\begin{figure}[!t]
\centering
 \includegraphics[trim=0 0 0 0cm, width=1.0\linewidth]{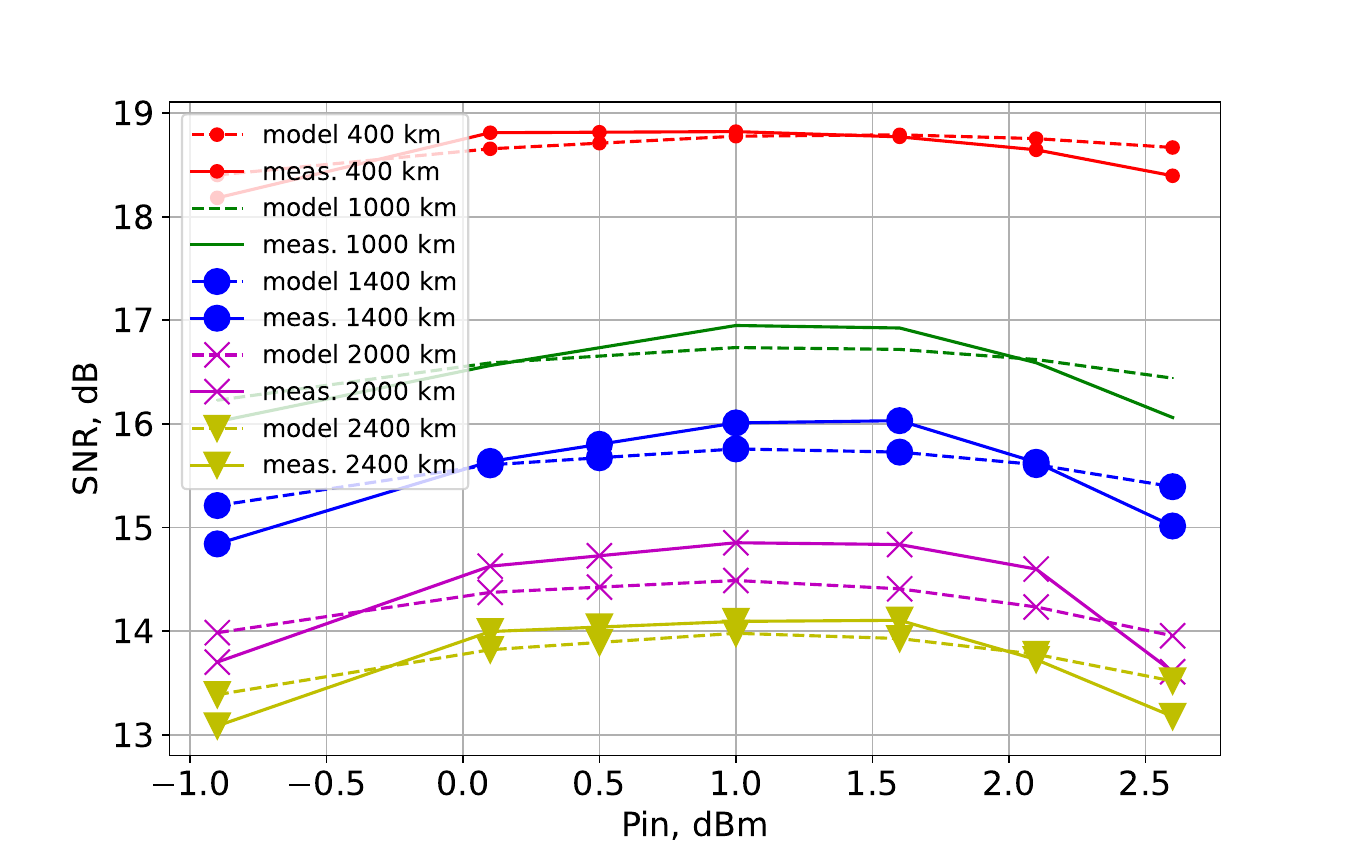}
 \caption{Experimental measurements of the received SNR for conventional 256QAM, together with the prediction of the model with fitted parameters.}
 \label{fig:fit}
\end{figure}

First, the received SNR is extracted for all studied distances and launch powers for 256QAM conventional constellation input. The quantization noise is maintained at 8 bits per real dimension, which is inline with the AWG and DSO specifications (observe, not effective number of bits, which as discussed in \sec{optimization} is taken into account in $SNR_{TRX}$). Then, the covariance matrix adaptation with evolutionary strategy (CMA-ES) gradient-free optimization algorithm is used to find the parameters which best explain the received effective SNR. The cost function for the fit is chosen to be the maximum error between model prediction and measurement across all distances and launch powers. The measurements, together with the model predictions when the fitted parameters are applied are given in \fig{fit} for distances of $400, 1000, 1400, 2000$ and $2400$ km. The fit results in a maximum error of $\approx 0.4$ dB, and the variations are attributed to the experimental measurement uncertainties, which are $\approx 0.2$ dB. The optimal parameters are found to be $NF=6.17$ dB, $SNR_{TRX}=20.78$ dB, $\alpha=0.183$ dB/km and $\gamma=0.986$ 1/W/km. Observe, this may be a local optimum of the objective function, which is highly non-convex. Additionally, not all components are explicitly accounted for with the model from \sec{optimization}, e.g. the connectors' losses. Regardless, these parameters are used to re-optimize the constellations for the subsequent measurements. As will be demonstrated in \sec{exp_results}, the obtained accuracy of the fit is sufficient to obtain similar gains to the idealized scenario in \sec{ideal_results}.   

\subsection{Channel fit and optimization in practical systems}

Similar characterizations can be applied prior to deployment of optical networks, or even after deployment without significant link downtime. One optimization run using the model described in \sec{optimization} takes approx. 2-3 minutes on a standard CPU and in most systems will be performed offline. Obtaining characterization points in terms of effective received SNR has negligible duration in a system using real-time transmission. In principle, the optimized constellations depend on the network load and other paramters, such as the launch power. However, we expect transferability to a high degree, with potentially only minor re-training (using the current constellation for initialization) necessary to obtain a new optimized constellation should the link path and network load configuration change. In cases where transmission parameters, such as the desired constellation can be software defined, the characterization method is therefore applicable to dynamic scenarios. Alternatively, a \textit{robust} set of constellations can be trained using e.g. variational methods \cite{OgnjenRobust}\cite{Karanov2019} or generalized structures \cite{RodeRobust} which are then hard-coded into the coherent transceivers and perform well on a wide variety of scenarios at the cost of some sub-optimality on individual scenarios.

\subsection{Zero-codeword argument}

As described in \sec{optimization}, even for fractional $n_d$, the optimization of the constellation shape is performed for $\lfloor n_d \rceil$. Fractional $n_d$'s are realized purely through tuning the true number $N_D$ to be mixed with the codeword. That means that the waveform distribution is different only up to an integer $n_d$ and modulation format size. In order to evaluate the possibility to tune the net data rate with a small step under constraints of experimental setup availability and finite time stability of the performance, only waveforms with an integer $n_d$ are transmitted (which include both optimized and conventional QAM), and with an FEC rate of $R=3/4$. Then, the zero codeword argument from \cite{Schmalen} is used in order to evaluate different coding schemes. With this method, the waveforms with an integer $n_d$ are processed with a fractional $n_d$ by the following. First, the true, noiseless transmitted QAM symbols are demapped using the labeling LUT. Then, the desired number of $N_D$ bits (which may be different than the true number for that trace) are discarded. The remaining bits are deinterleaved. The resulting bit sequences are treated as scrambling sequences for the all-zero codeword, which is then assumed to have been generated by the target FEC (which may also be different than the original FEC for that trace). At the receiver, the LLRs are obtained from the noisy symbols, the desired number of bits $N_D$ are discarded, and the desired number of coded bits $N$ are deinterleaved. The scrambling sequences are applied to the deinterleaved noisy LLRs, the desired FEC decoder is run, and the BER is estimated against the all-zero codeword.

\section{Experimental results}
\label{sec:exp_results}

\begin{figure*}[!t]
\centering
\subfigure{
 \includegraphics[trim=0 0 0 0cm, width=0.31\linewidth]{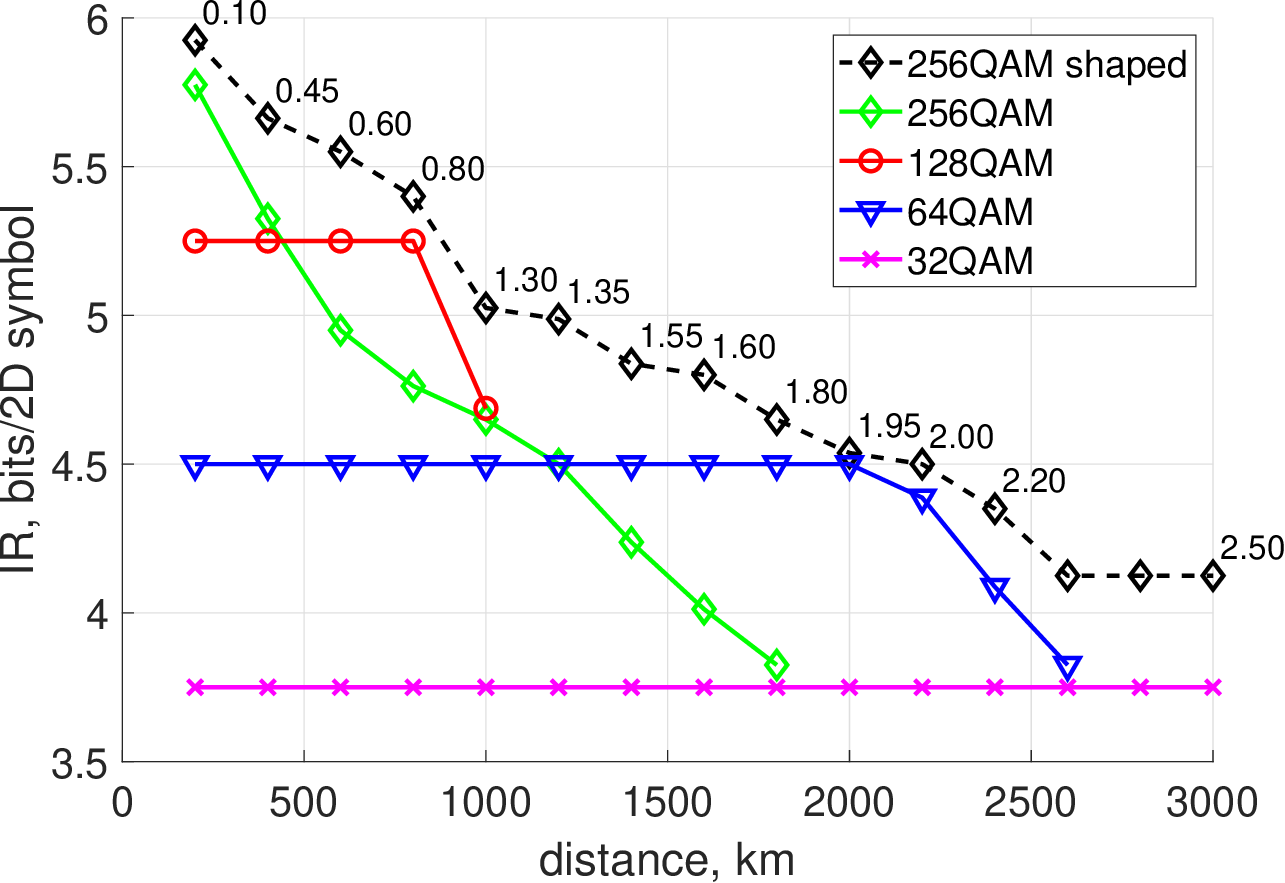}}
 \subfigure{
 \includegraphics[trim=0 0 0 0cm, width=0.31\linewidth]{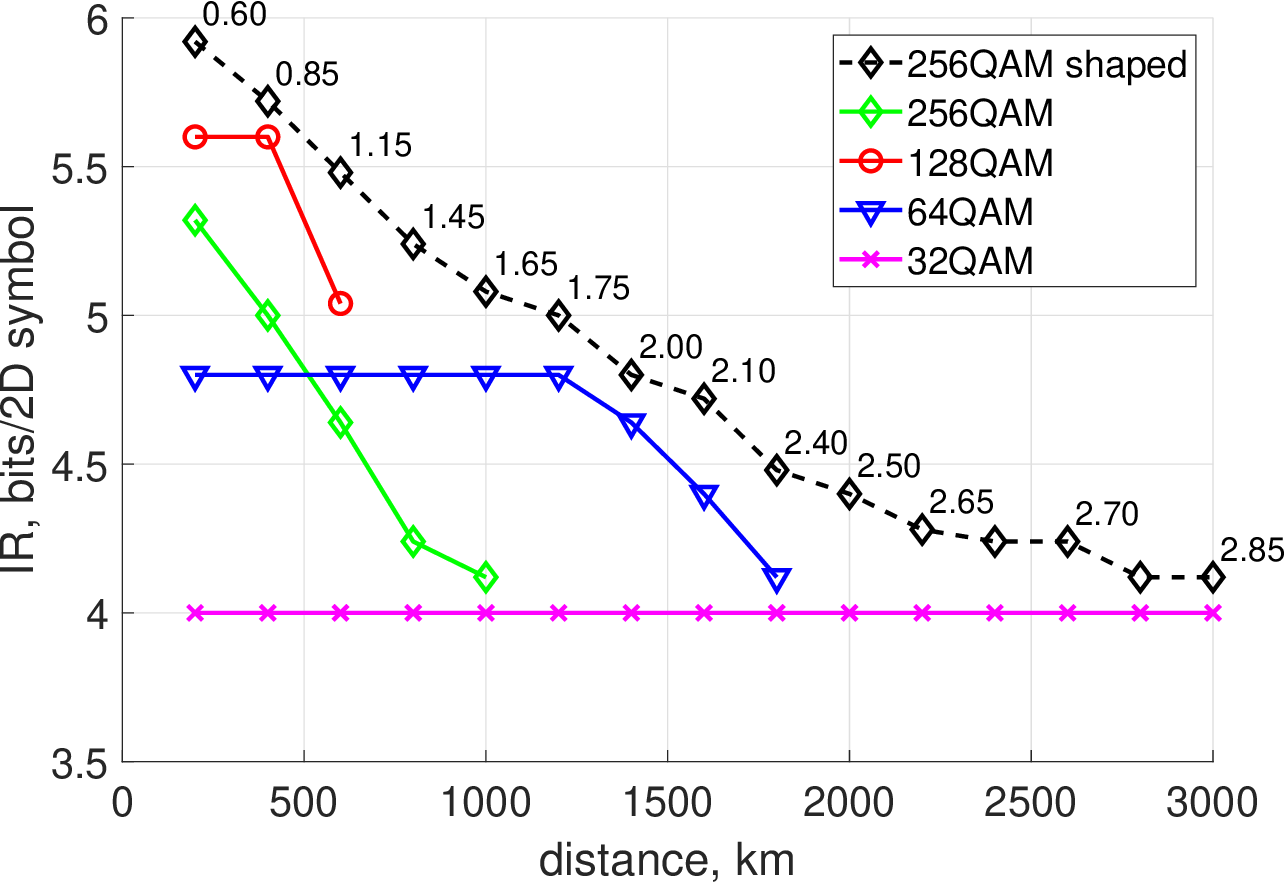}}
 \subfigure{
 \includegraphics[trim=0 0 0 0cm, width=0.31\linewidth]{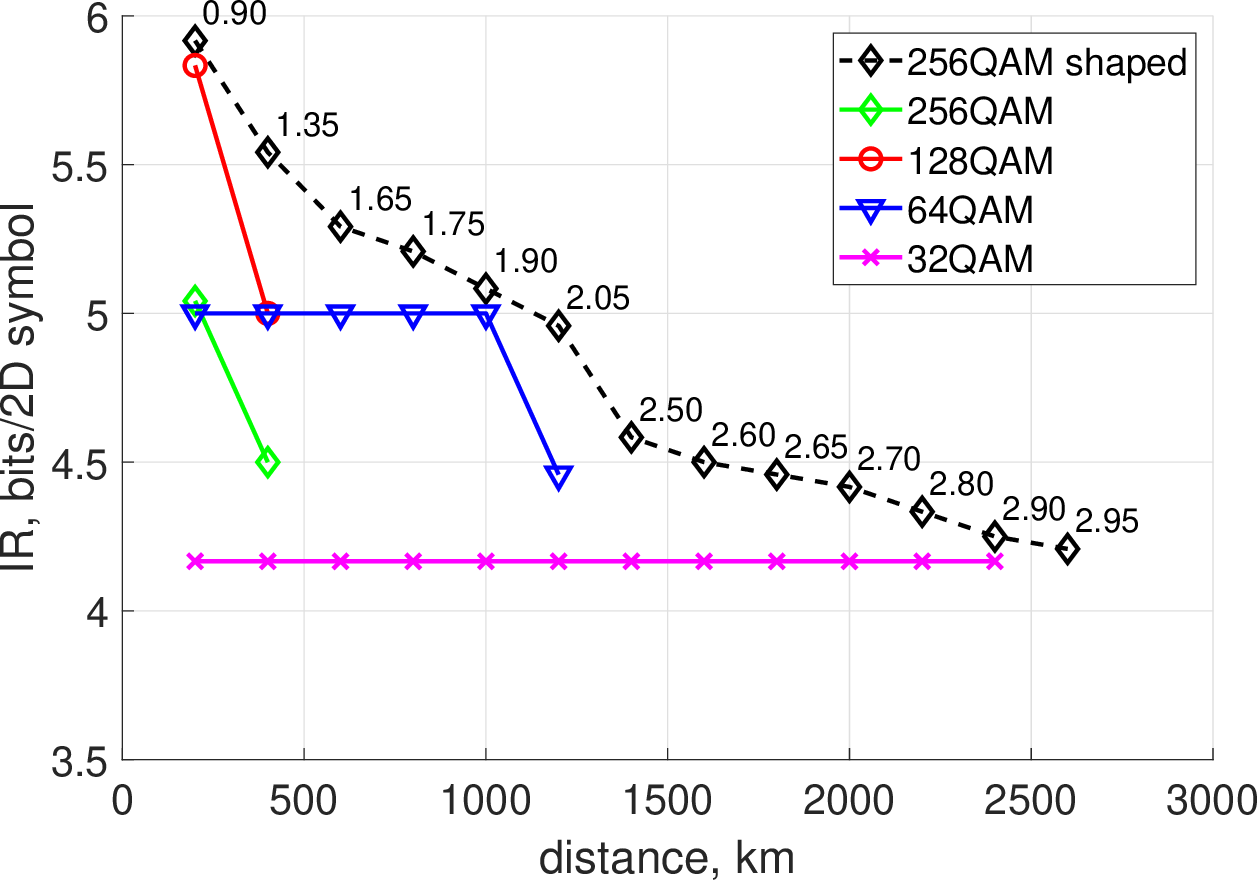}}
 \\
 \subfigure{
 \includegraphics[trim=0 0 0 0cm, width=0.31\linewidth]{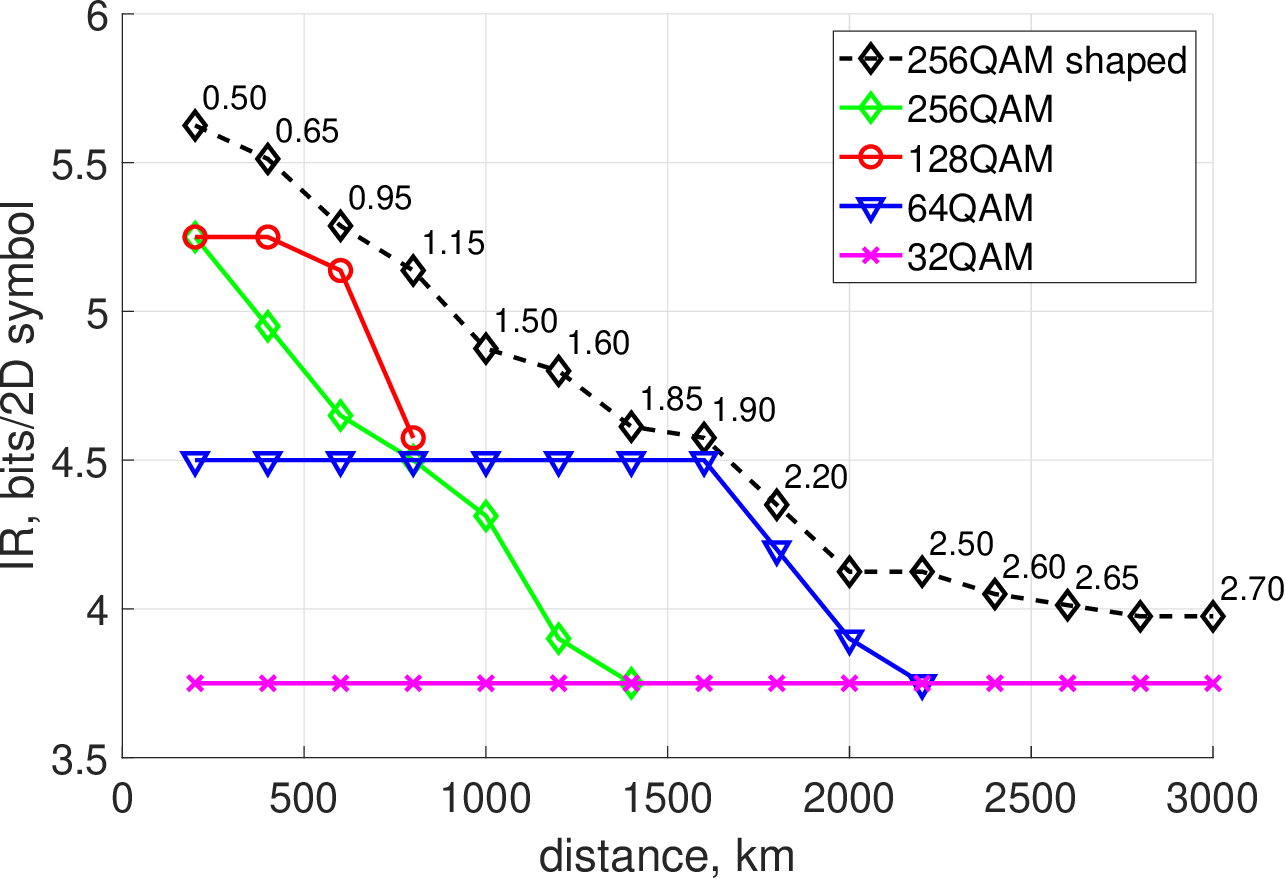}}
 \subfigure{
 \includegraphics[trim=0 0 0 0cm, width=0.31\linewidth]{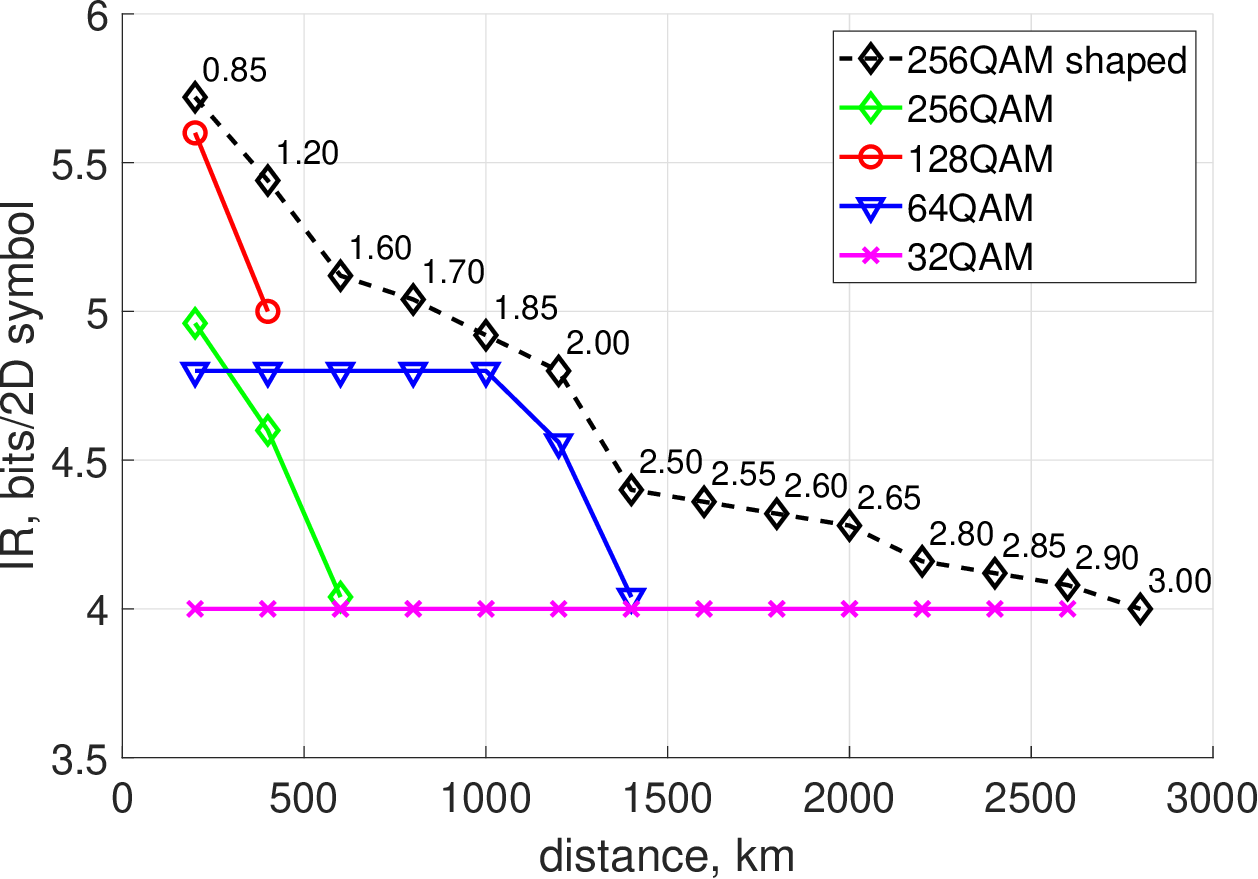}}
 \subfigure{
 \includegraphics[trim=0 0 0 0cm, width=0.31\linewidth]{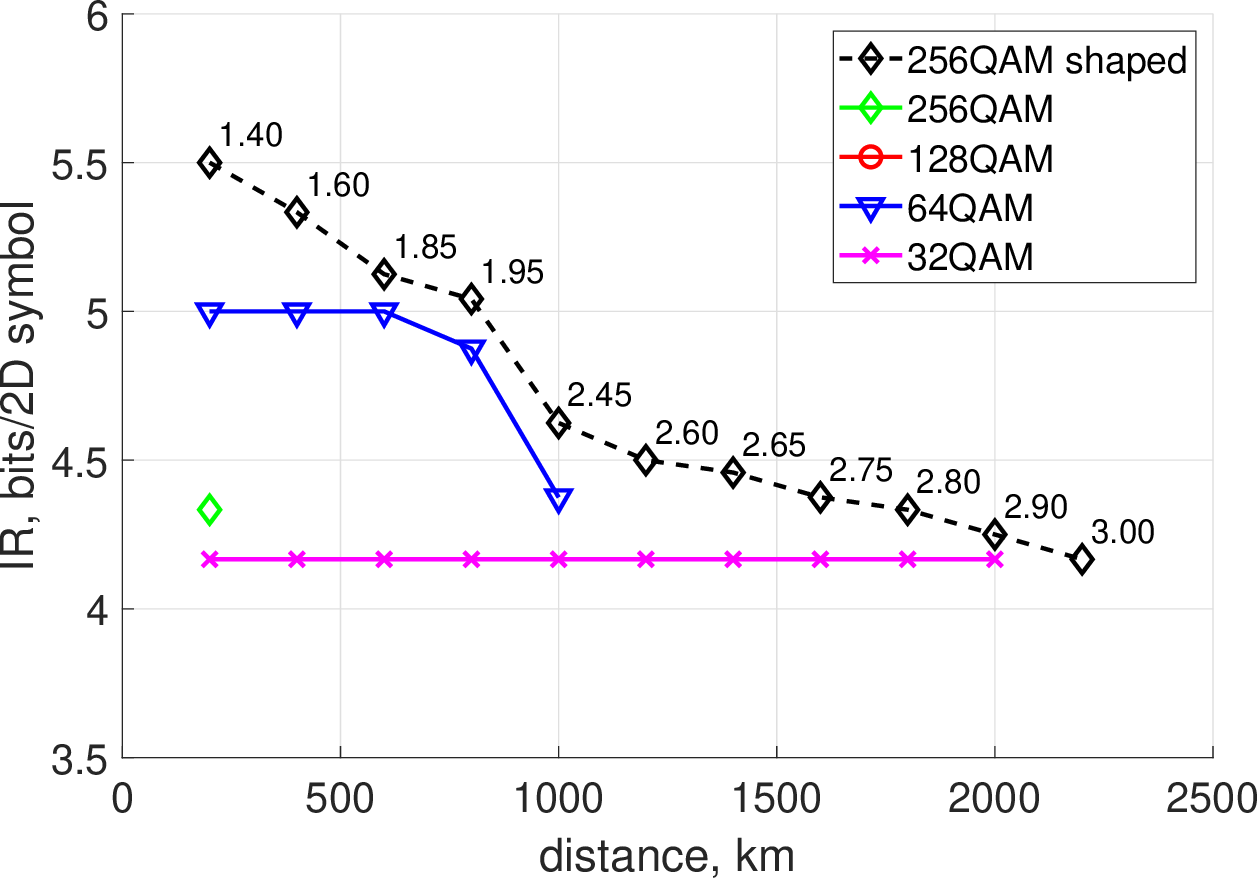}}
 \caption{Experimental results for maximum achievable distance with the target rate. All constellation consider MTOM \textbf{Top row:} assuming ideal FEC based on maximum distance where AIR $>$ IR; \textbf{Bottom row:} with LDPC FEC, where $BER < 10^{-5}$ at the target IR. Results for 33\%, 25\% and 20\% FEC overhead in the first, second and third column, respectively. The number $n_d$ is annotated on the performance curve of the shaped constellations.}
 \label{fig:exp_results_GCS}
\end{figure*}

Similar to the AWGN case, the numbers $n_d$ which are evaluated are swept in the range $n_d \in \left[0, \min\{3, m-5\}\right]$ with a step of 0.05. Geometric shaping is only applied to 256QAM. In each case, 3 independent traces are captured with more than 20 LDPC codewords per trace, which in the lowest-rate scenario corresponds to 2.916.000 information bits. In \fig{exp_results_GCS}, the maximum error-free distance is reported. On the top row, the idealized case is shown, where a rate is considered achieved if the AIR is above it. On the bottom row, the practical FEC-based case is shown, where an error-free performance is assumed when the BER after LDPC decoding is $<10^{-5}$. 

\begin{figure}[!t]
\centering
\subfigure{
 \includegraphics[trim=0 0 0 0cm, width=0.31\linewidth]{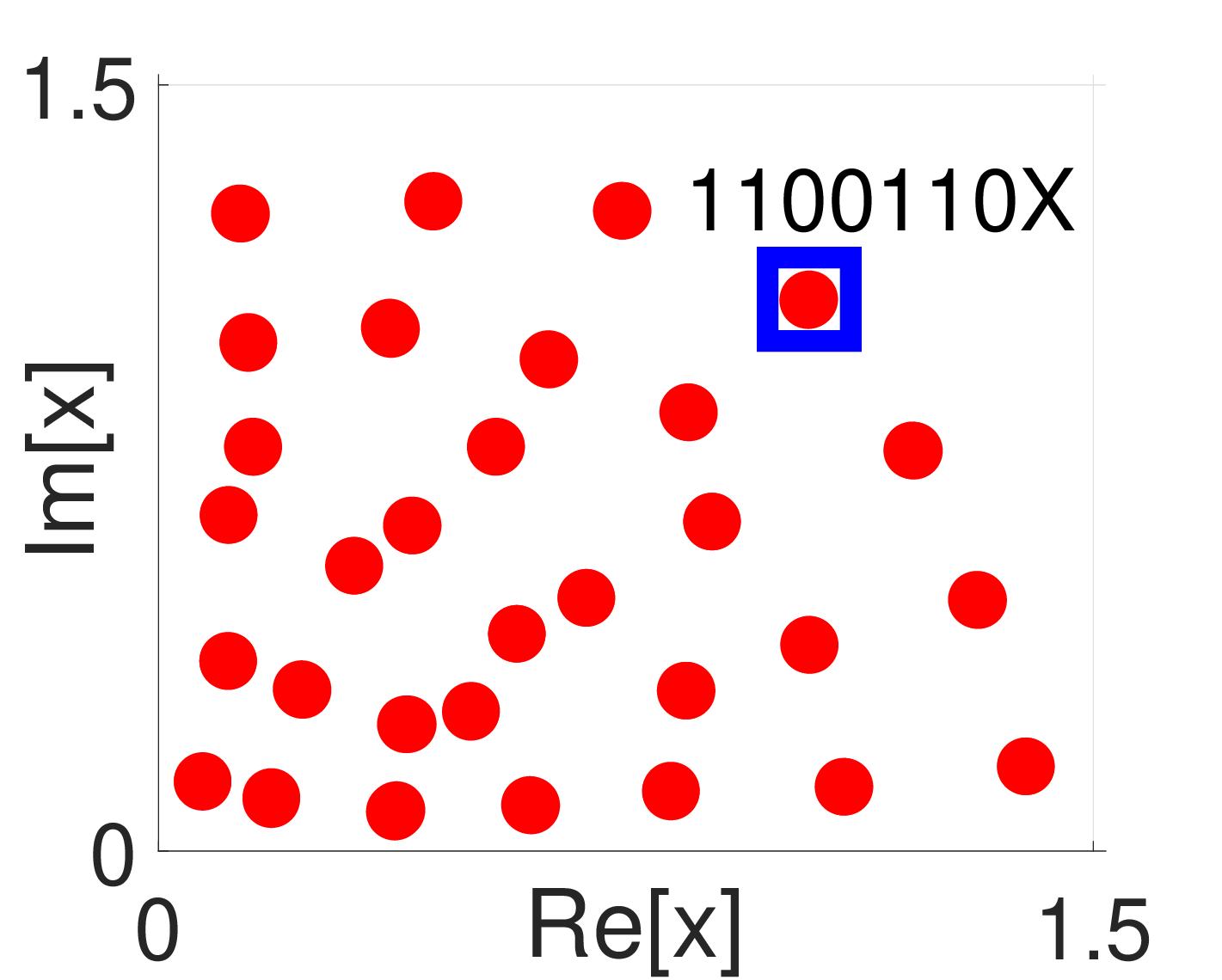}}
 \subfigure{
 \includegraphics[trim=0 0 0 0cm, width=0.31\linewidth]{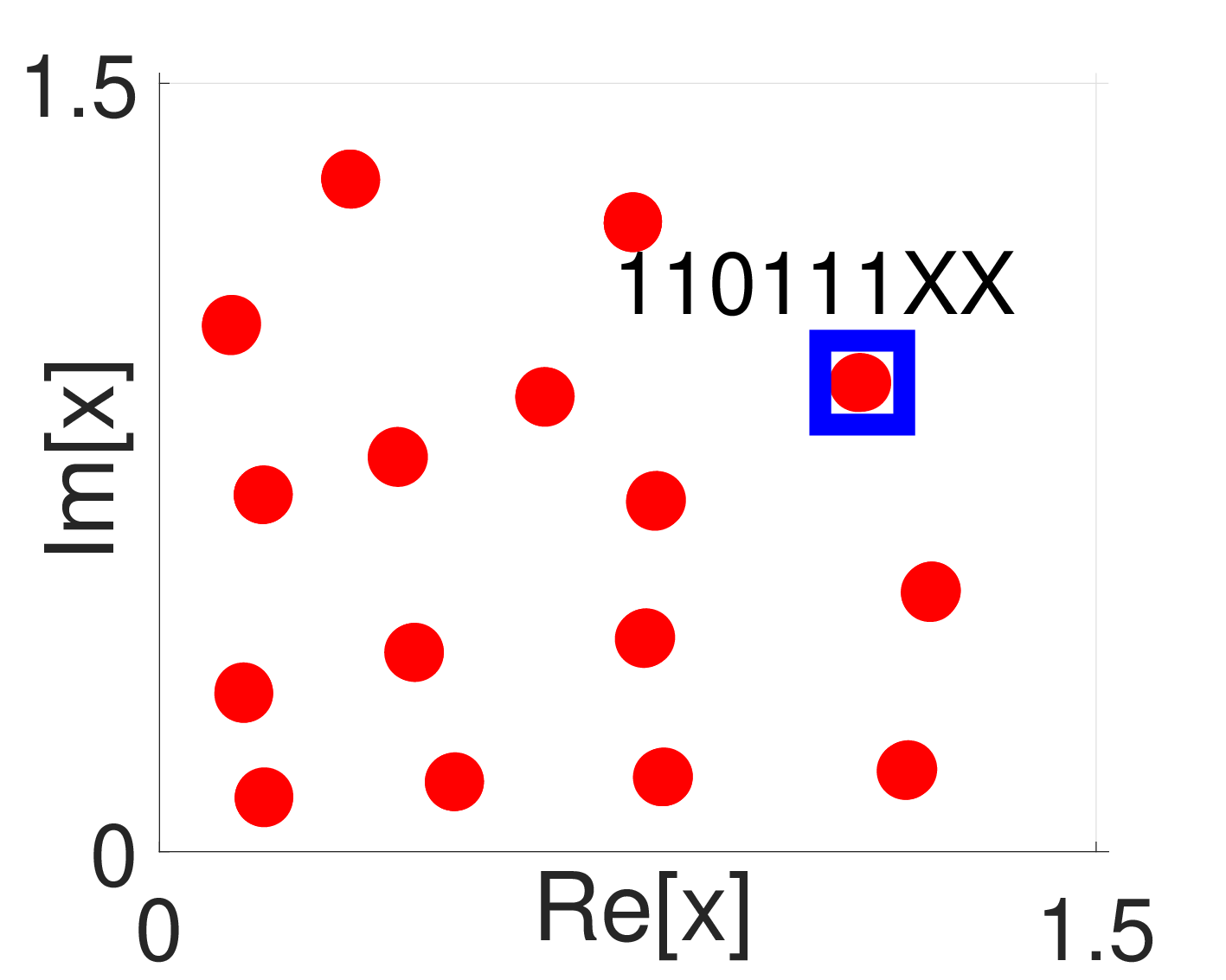}}
 \subfigure{
 \includegraphics[trim=0 0 0 0cm, width=0.31\linewidth]{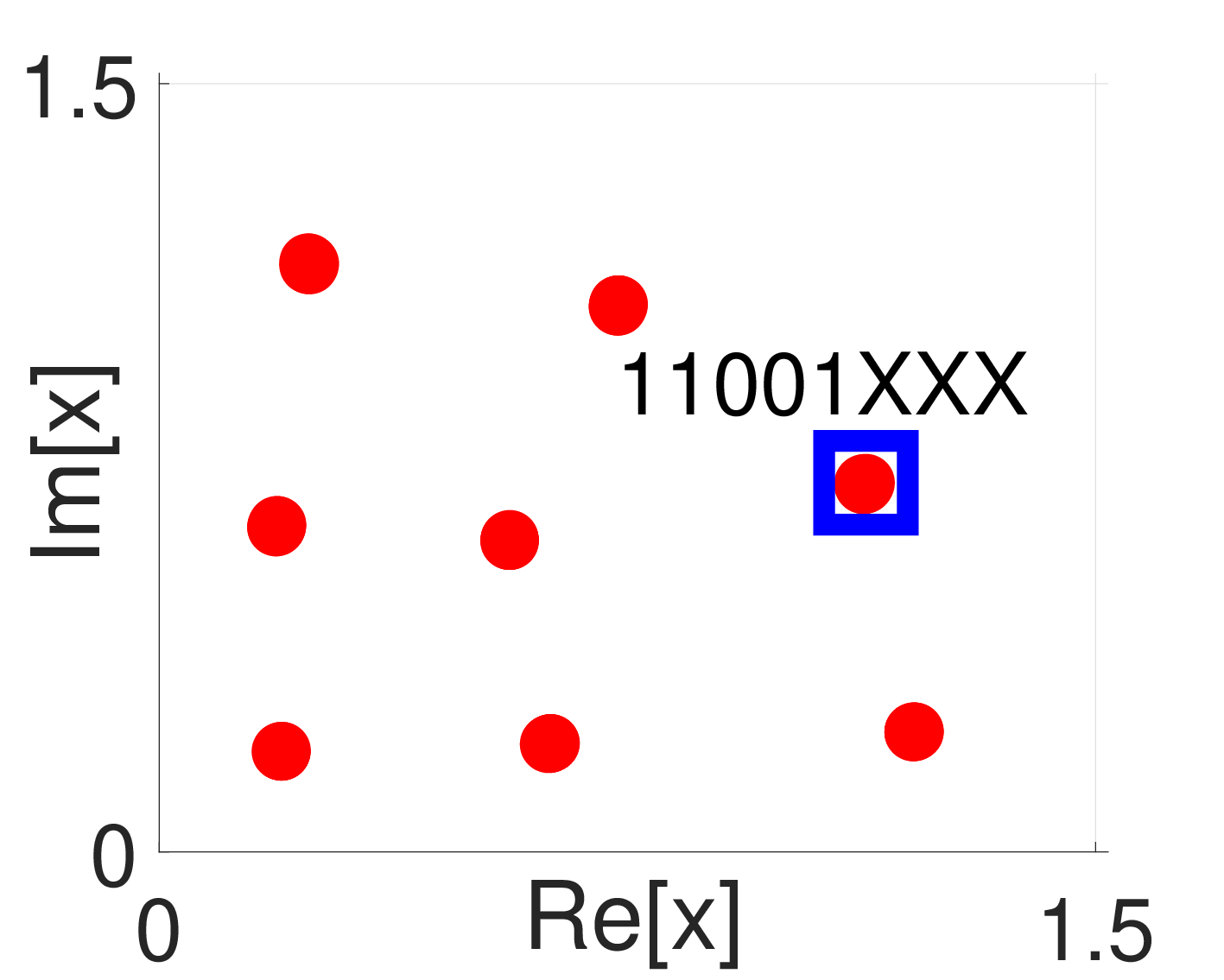}}
 \caption{Example constellations $\X_{red}$ (from left to right): $n_d=1$ optimized for 400 km; $n_d=2$ optimized for 1400 km; and $n_d=3$ optimized for 2400 km. The label of one set of nearly-identical points is exemplified, where 'X' illustrates that the corresponding bit may take any value. }
 \label{fig:example_constellations}
\end{figure}

We begin the analysis of the performance with the case of $R=3/4$, represented in the left two subfigures of \fig{exp_results_GCS}. Uniform QAMs without MTOM ($n_d=0$) achieve an FEC-based error-free performance for up to 400 km, 1600 km and 3000 km (maximum length studied) for 128QAM, 64QAM, and 32QAM, respectively. Error-free performance could not be achieved with 256QAM with $n_d=0$. As also demonstrated by the NLIN simulations, unshaped QAM with MTOM can be utilized with a reasonable performance penalty for up to $n_d<1$, after which the MI on the data bits is insufficient to ensure error-free decoding. Shaping through optimization is realized by merging the points with identical data-carrying part of the label, significantly boosting the MI of those bits. Examples of such constellations are shown in \fig{example_constellations} in the case of $n_d=1$ for 400 km, $n_d=2$ for 1400 km and $n_d=3$ for 2400 km. Each `point' of the constellations is in fact the superposition of 2, 4 and 8 points, respectively, which have nearly identical coordinates. The leading part of the label of these points is identical and is exemplified for 1 of the points from each constellation. The `dummy' bits take any value. 

The gain of the proposed scheme w.r.t. conventional, unshaped, BICM BRGC QAM can be estimated to be up to 0.79 bits/2D e.g. for 600 km, where the the conventional BICM needs to revert to 64QAM with a large margin and a net data rate of 4.5 bits/2D symbol, whereas the proposed scheme operates at 256QAM with $n_d=0.95$ for a net data rate of 5.29 bits/2D symbol. Similar gains follow for each distance where the conventional BICM needs to reduce the modulation format size because the proposed scheme is able to gradually reduce the rate, instead of by $R$ as the conventional scheme. At the same rate as conventional BICM, the proposed system achieves a shaping gain of up to 2 spans, measured at the rate just before the modulation format size needs to be reduced.

Similar analysis follows for the other FEC cases, with the exception that high-order conventional BRGC QAM generally is not supported by such FEC because in such cases, $n_d$ needs to be large in order to operate at the optimal rate supported by the effective received SNR. On the other hand, the proposed scheme achieves similar performance regardless of the FEC rate. At lower $R$, lower $n_d$ is selected, whereas at higher $R$, higher $n_d$ is selected with an overall similar performance in the two cases. 

The proposed scheme appears to exhibit a penalty w.r.t. the overall trend for cases of high $R$ and where $n_d > \lfloor n_d \rceil$. At such operating points, a lot of non-zero MI is wasted on dummy bits, which at the same time prevents increasing the MI of the data-carrying bits due to insufficient merging of constellation points. We experimented also with selecting $\lceil n_d \rceil$ for optimization, which counters this problem, but on the other hand introduces too many erasures at data-carrying label positions. This drawback of the proposed method will also be seen in the \sec{PAS}, where comparison to PAS is made, and is a very relevant topic for future research. It can be approached by the above-mentioned erasure-aware FEC and TH modulation.

\begin{figure}[!t]
\centering
 \includegraphics[trim=0 0 0 0cm, width=1.0\linewidth]{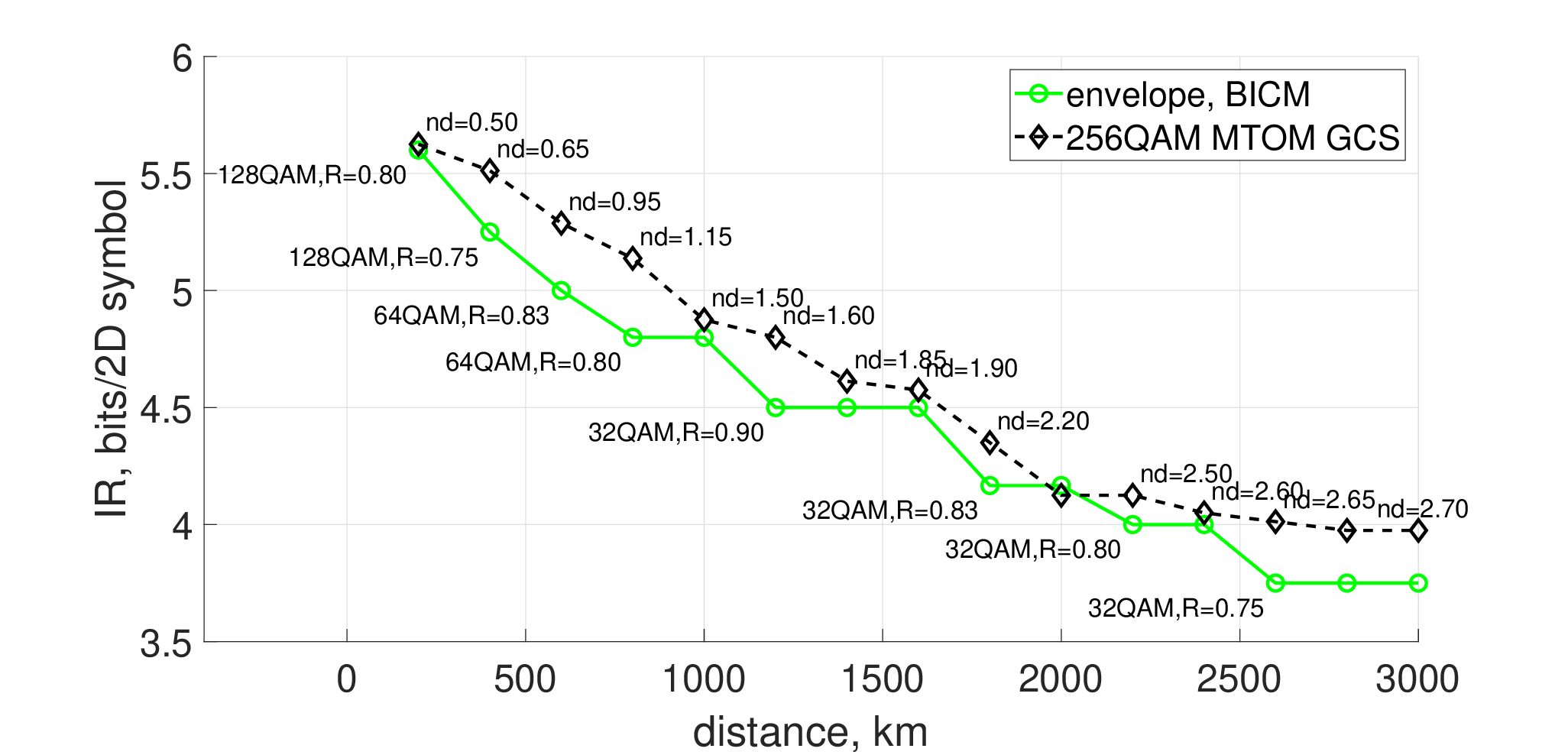}
 \caption{Comparison between the proposed method and conventional BICM which can achieve rate adaptation by flexibility in the FEC rate and modulation format size.}
 \label{fig:comparison_RA_BICM}
\end{figure}

For completeness, the proposed scheme is also compared to a BICM for which the rate is adopted by using flexible-rate FEC and flexible modulation format size. The family of DVBS-2 codes are standardized for rates $1/4, 1/3, 2/5, 1/2, 3/5, 2/3, 3/4, 4/5, 5/6, 8/9$ and $9/10$. All combinations of those rates and the studied modulation formats 32QAM, 64QAM, 128QAM and 256QAM are evaluated, the combination with a highest achieved error-free rate after LDPC deocding is selected at each distance and comapred to the proposed scheme with $R=3/4$. The comparison is given in \fig{comparison_RA_BICM}, where the optimal BICM combination is annotated on the left, and the selected $n_d$ for the proposed scheme on the right. Shaping gain of up to $\approx 0.3$ bits/2D symbol is achieved with the proposed scheme w.r.t. the highly impractical rate adaptation method.   

\subsection{Comparison to PAS}
\label{sec:PAS}

\begin{figure*}[!t]
\centering
\subfigure{
 \includegraphics[trim=0 0 0 0cm, width=0.31\linewidth]{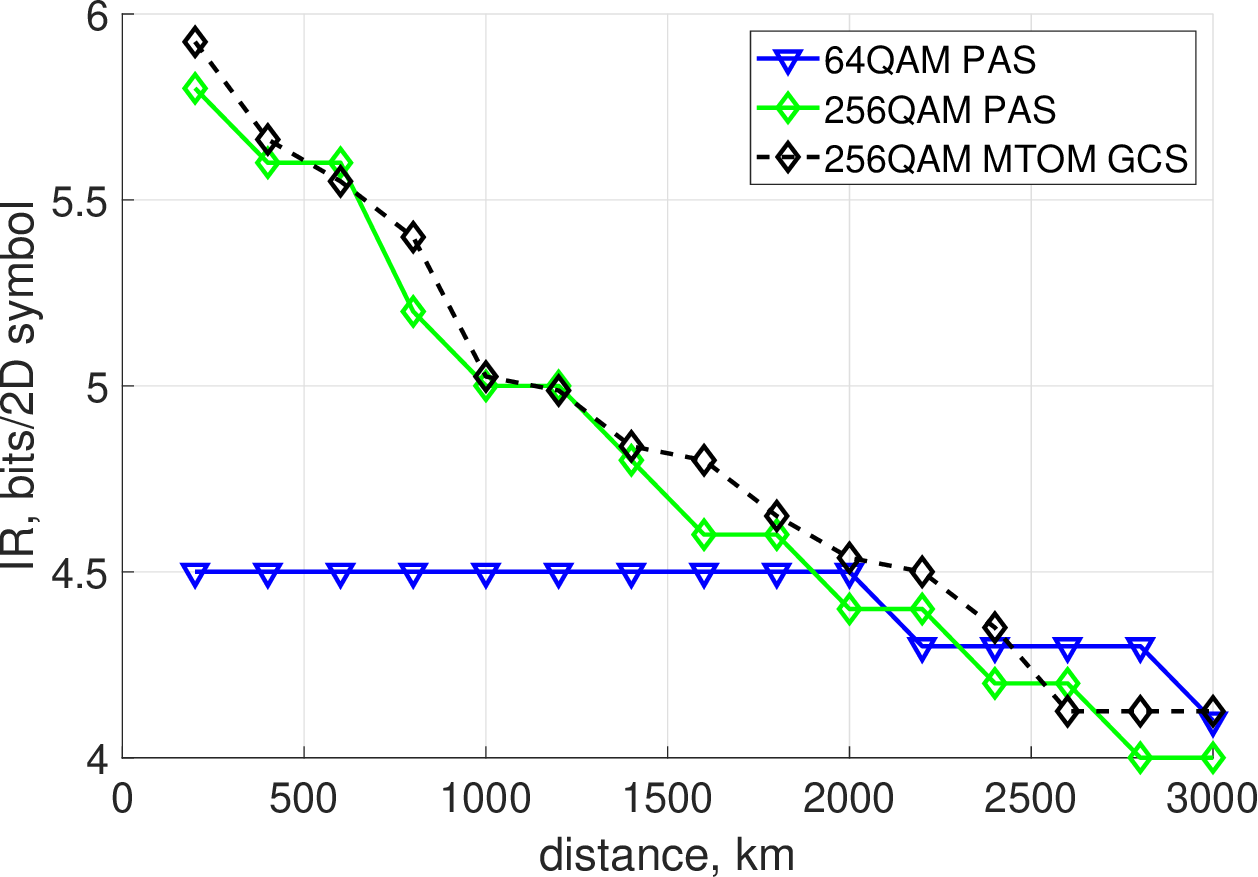}}
 \subfigure{
 \includegraphics[trim=0 0 0 0cm, width=0.31\linewidth]{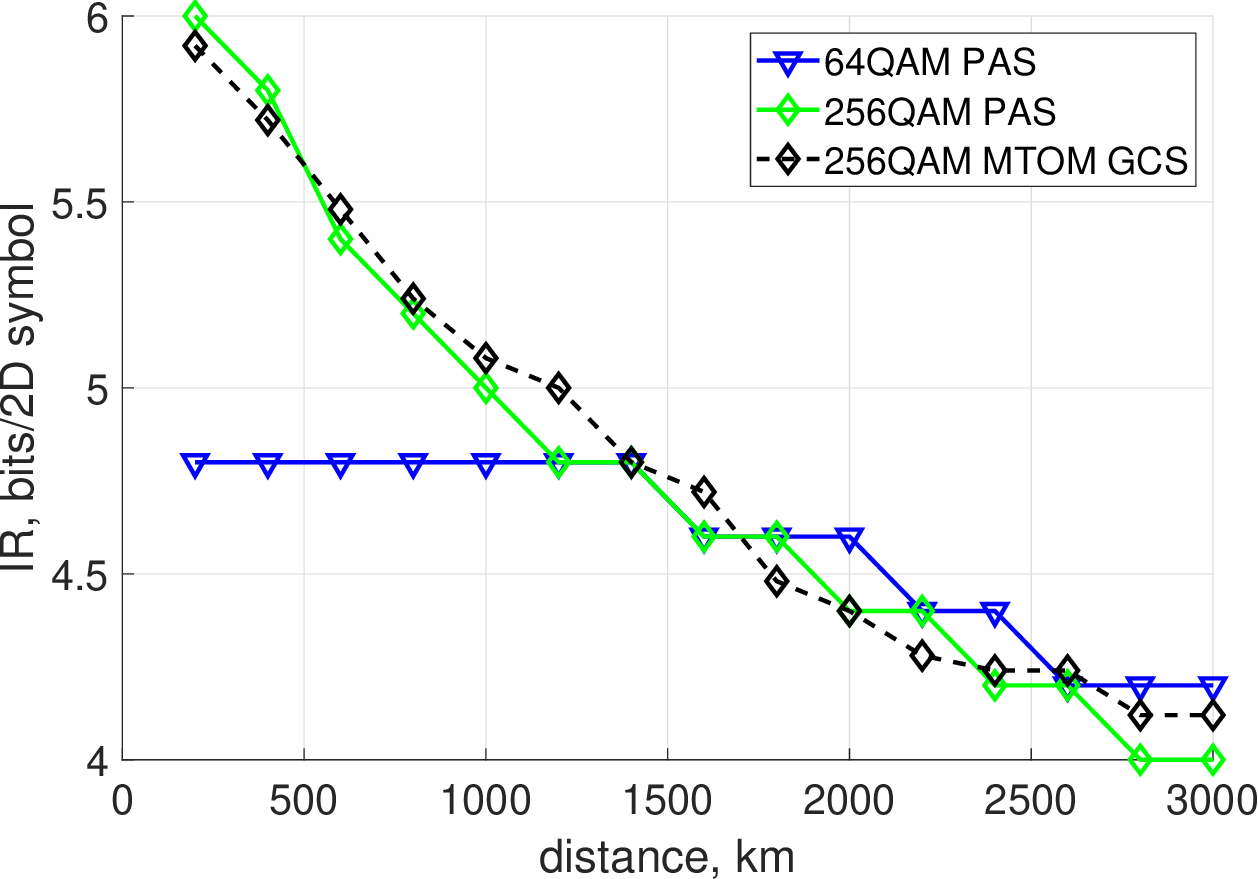}}
 \subfigure{
 \includegraphics[trim=0 0 0 0cm, width=0.31\linewidth]{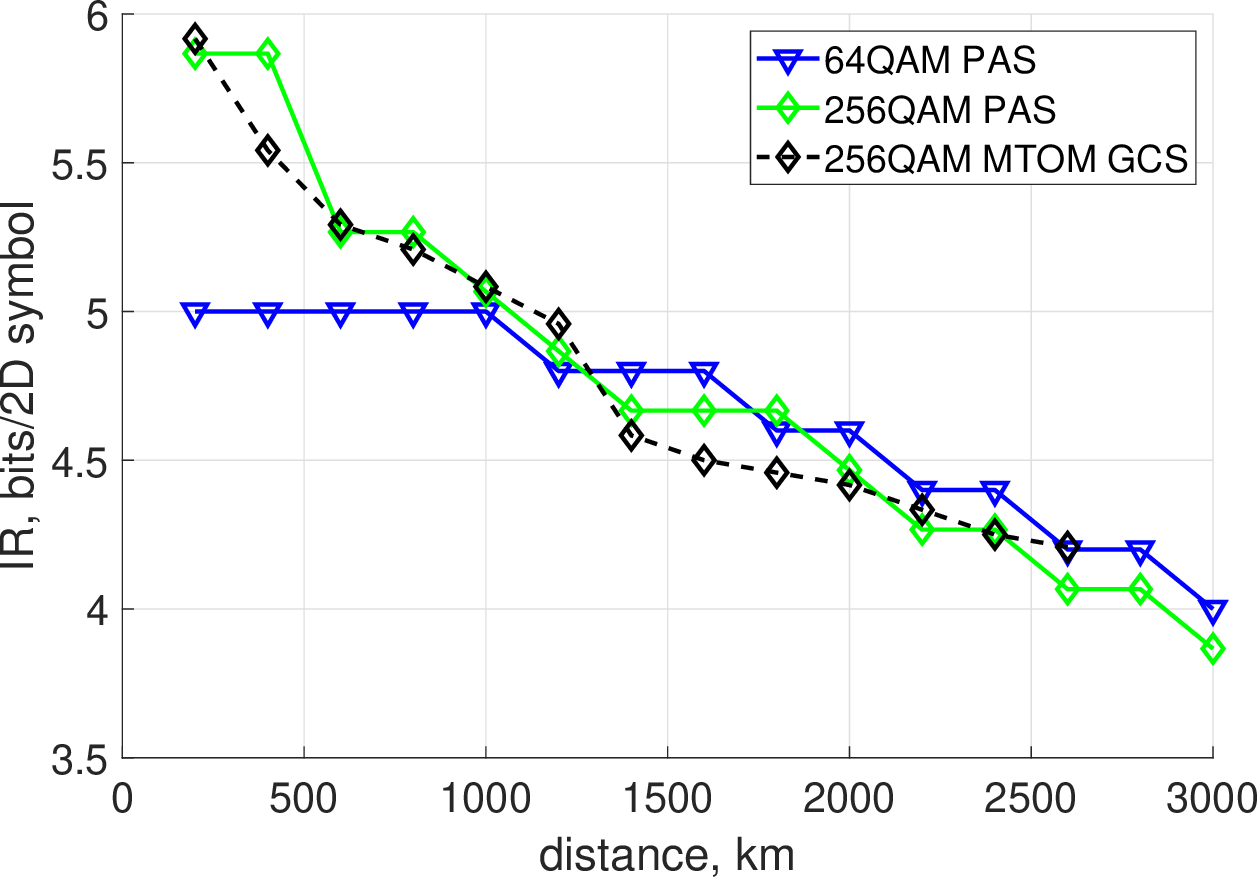}}
 \\
 \subfigure{
 \includegraphics[trim=0 0 0 0cm, width=0.31\linewidth]{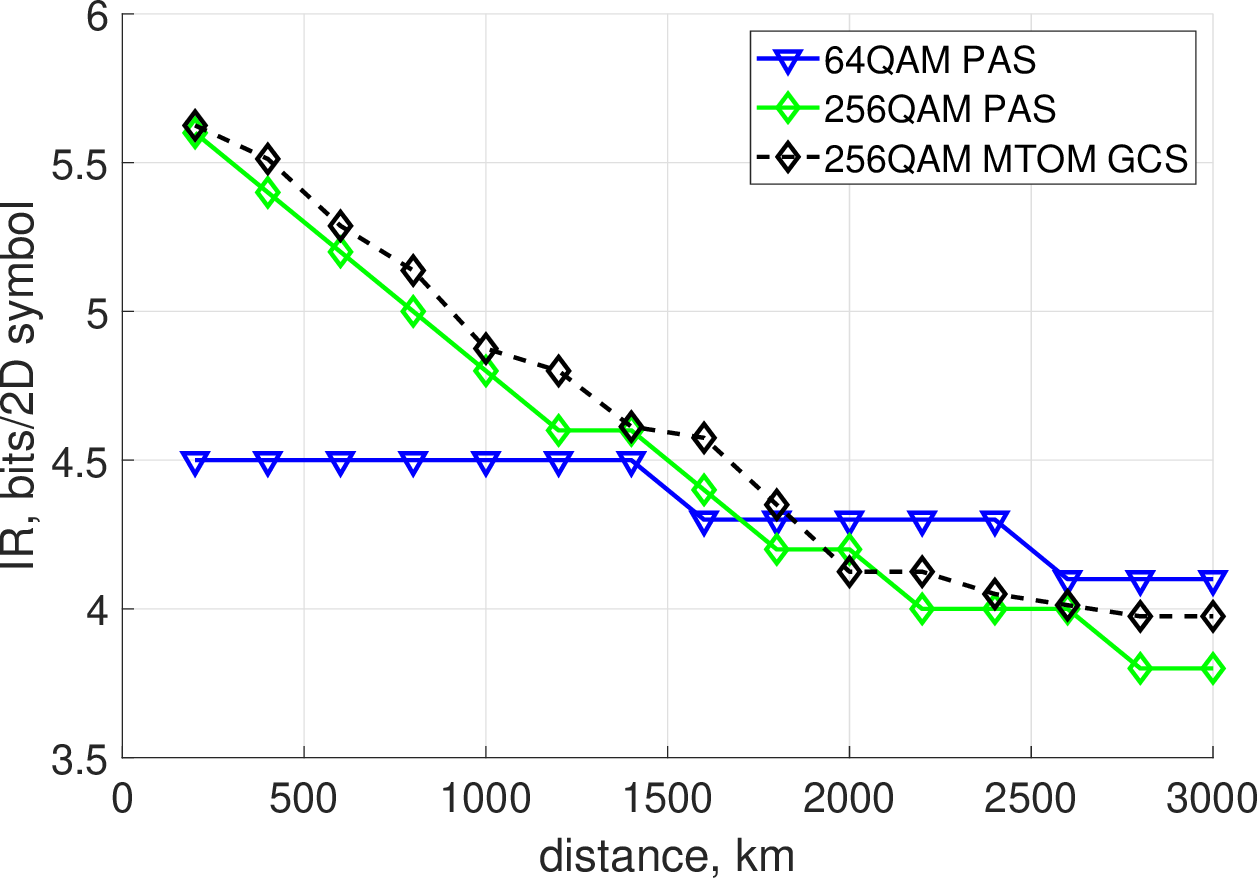}}
 \subfigure{
 \includegraphics[trim=0 0 0 0cm, width=0.31\linewidth]{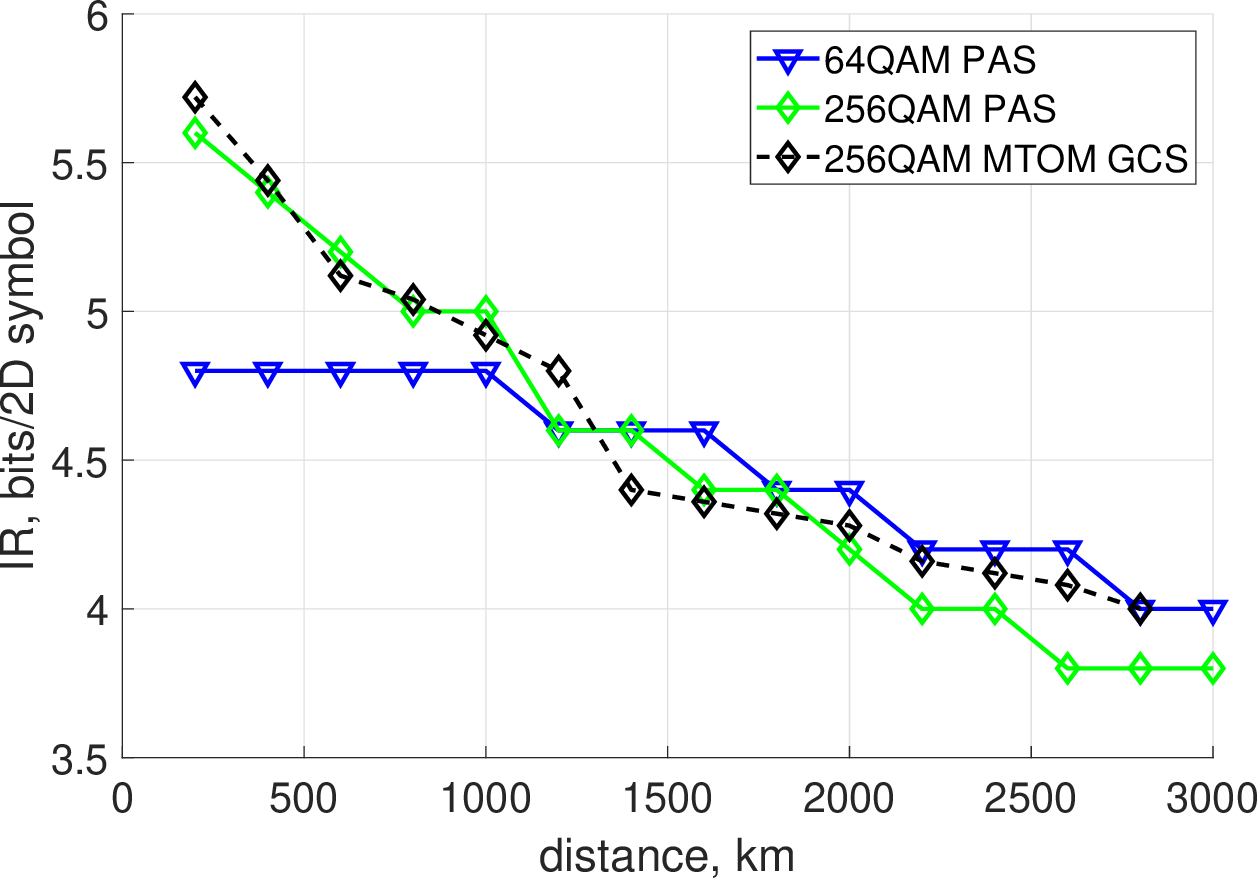}}
 \subfigure{
 \includegraphics[trim=0 0 0 0cm, width=0.31\linewidth]{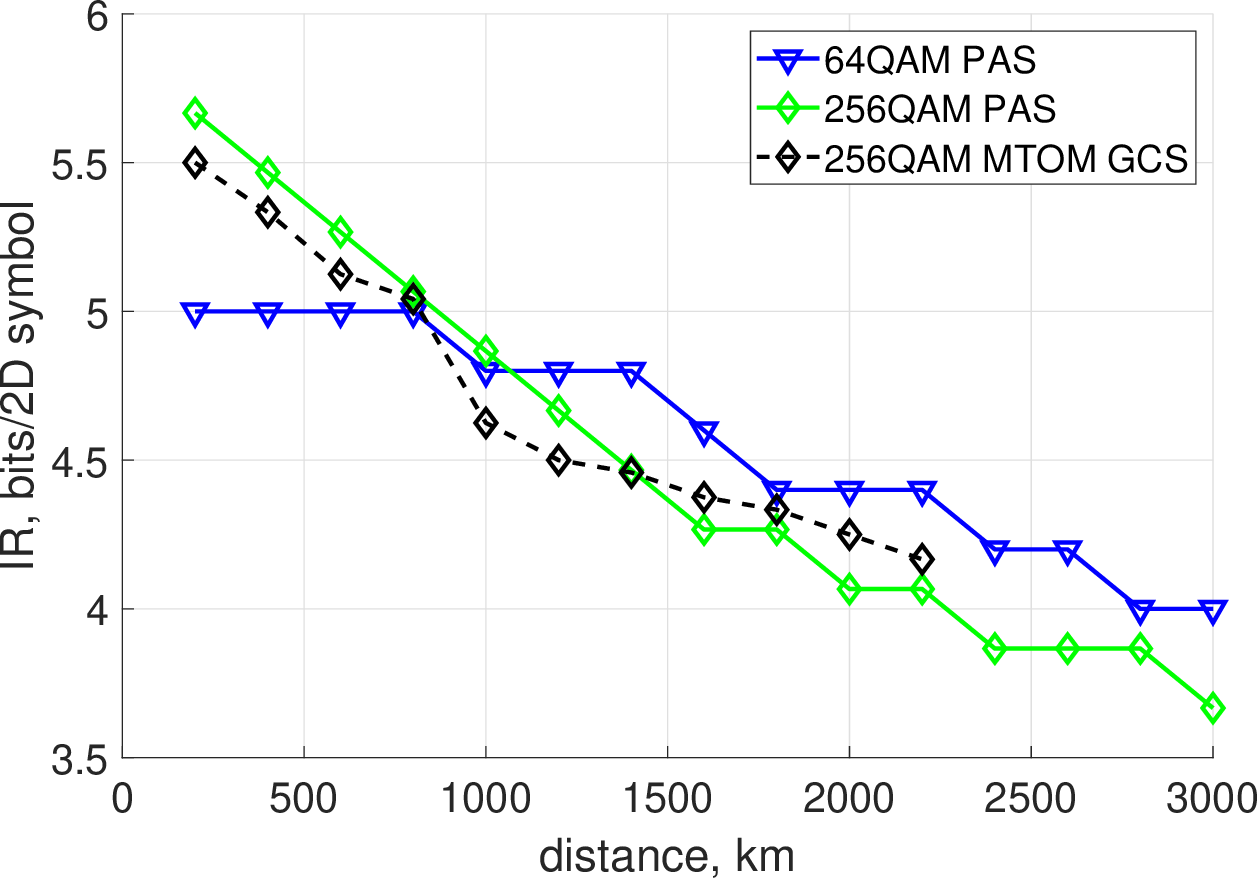}}
 \caption{Experimental results for maximum achievable distance with the target rate using PAS as described in \sec{PAS}, compared to the proposed MTOM GCS. \textbf{Top row:} assuming ideal FEC based on maximum distance where AIR $>$ IR; \textbf{Bottom row:} with LDPC FEC, where $BER < 10^{-5}$ at the target IR. Results for 33\%, 25\% and 20\% FEC overhead in the first, second and third column, respectively.}
 \label{fig:exp_results_PAS}
\end{figure*}

The PAS scheme from \cite{Bocherer} is also studied on the same experimental test bed for completeness and comparison to the proposed MTOM GCS. PAS is realized similar to the AWGN case using the same FEC as above, and the MB PMF obtained using a CCDM \cite{Schulte}.  Similarly to the GCS case, waveforms generated with a FEC rate of $R=3/4$ are transmitted and captured for each $\H(X)$, and the performance with other FEC rates is evaluated using the zero-codeword method \cite{Schmalen}.

In \fig{exp_results_PAS}, the performance of PAS is given in terms of the maximum error-free distance for both the idealized case (top row) where the \textit{normalized} AIR $>$ IR, and the practical FEC case (bottom row), where the error-free rate is extracted as explained above from the BER after LDPC decoding. Similar to the proposed GCS case, FEC rates of $R=3/4$, $R=4/5$ and $R=5/6$ are studied (left, middle and right column, respectively), with the same LDPC FEC from the DVBS-2 standard. For comparison, the performance of the MTOM GCS is also given. 

The proposed scheme generally achieves similar performance to PAS with 256QAM at the high-to-moderate rates, and is slightly better than 256QAM PAS at the lower rates. However, unlike the idealized scenarios in previous sections, PAS with 64QAM appears to be slightly better than PAS with 256QAM and the proposed scheme. This is especially the case in the operating points with $n_d > \lfloor n_d \rceil$ mentioned above, where the proposed scheme experiences performance deficiencies. In general, 64QAM PAS is a very robust modulation format, but as just mentioned, does not support very high spectral efficiencies. The average gain of PAS w.r.t. the proposed system over all distances if the modulation format is allowed to change between 256QAM and 64QAM is 0.004, 0.045, and 0.184 bits/2D symbol for $R=3/4$, $R=4/5$ and $R=5/6$, respectively. The gain of PAS when restricted to 256QAM w.r.t. the proposed system is on the other hand -0.095, -0.054, and 0.038 bits/2D symbol for $R=3/4$, $R=4/5$ and $R=5/6$, respectively.

\section{Additional discussions}
\label{sec:discussion}
The CCDM used here to realize PAS is not the most practical matcher due to the required long length. There has been a variety of matcher architectures which improve on the complexity and parallelization capabilities of DMs (a summary can be found in e.g. \cite{Stella}). Furthermore, the MB PMF is also not the most suitable choice for PAS when applied to the nonlinear fiber channel. It should be noted that other functional PMF forms are also under investigation in the research community and have shown some benefit in terms of AIR, e.g. the super-Gaussian PMF family \cite{Infinera}. On the other hand, the CCDM of long length studied in this paper represents an idealized minimal rate loss but at an impractical complexity. The choice of PAS with CCDM and the MB PMF is driven by the desire to compare GCS to a more widely accepted high-performance architecture. A thorough optimization of the PAS scheme can potentially reveal additional gains to the ones demonstrated here, but is beyond the scope of this paper. 

We also note that erasures/punctures resulting from MTOM may have a detrimental effect on the prediction of the performance of the FEC. Moreover, for very low error rate operation, the proposed scheme requires the corresponding simulation to confirm that no error-floor appears. As mentioned above, a potential error-floor may be addressed either by an erasure-aware FEC, or by adopting the TH variant of the scheme.

\section{Conclusion}
\label{sec:conclusion}
In this work, a geometric constellation shaping scheme was proposed and experimentally demonstrated. The scheme allows to tune the target net data rate with an arbitrarily small step, which is imperative for future heterogenous network aiming at operating at minimal margin. The scheme may be interpreted by a many-to-one mapping in that neither the modulation format size nor the underlying forward error correction engine need to change in order to change the rate. This results in a very simple ASIC architecture, which is furthermore fully backwards compatible with conventional bit-interleaved coded modulation (BICM) implemented in current state of the art optical coherent transceivers. The added rate adaptivity functionality comes at a very minimal expense w.r.t. the state of the art BICM by only requiring additional labeling look-up tables (LUTs) for constellation mapping. The LUTs can be obtained offline using automatic differentiation using a pre-characterized surrogate channel model, or in semi-real time for arbitrary configuration of the network load. Furthermore, in the nonlinear fiber channel case, the proposed scheme exhibits similar performance to the popular probabilistic amplitude shaping (PAS) scheme without the requirement of PAS for a distribution matcher. The proposed scheme is an excellent candidate for the next generation of flexible coherent optical transceivers. 

\section*{Acknowledgment}
This work was supported by the Danish National Research Foundation Centre of Excellence SPOC, grant no. DNRF123, and Villum Foundation Young Investigator project OPTIC-AI, grant no. VIL29334. We would like to thank OFS Denmark for providing the SCUBA fiber used in the experiment.

\bibliographystyle{IEEEtran}
\bibliography{references}

\end{document}